\documentclass[aps, prd, amsmath, floats, floatfix, twocolumn,superscriptaddress, nofootinbib, showpacs]{revtex4}

\usepackage{graphicx}
\usepackage{amsmath,amssymb}
\usepackage{amsfonts}
\usepackage{xspace} 
\usepackage[usenames]{color}
\usepackage{dcolumn}
\usepackage{bm}

\usepackage{ulem}
\newcommand{\Caltech}{\affiliation{Theoretical Astrophysics 130-33,
    California Institute of Technology, Pasadena, CA 91125}}
\newcommand{\Cornell}{\affiliation{Center for Radiophysics and Space
    Research, Cornell University, Ithaca, New York, 14853}}
\newcommand{\Maryland}{\affiliation{Maryland Center for Fundamental
    Physics, Department of Physics, University of Maryland, College
    Park, MD 20742}}

\newcommand{\OrbitalPhase}{\Phi}

\newcommand{\OrbitalFreq}{\Omega}

\definecolor {darkgreen}{rgb}{0.2,0.7,0.2}

\newcommand{\pade}{Pad\'{e}\xspace}

\begin{document}

\title{Effective-one-body waveforms calibrated to numerical relativity simulations: coalescence of non-spinning, equal-mass black holes}

\author{Alessandra Buonanno} \Maryland %
\author{Yi Pan} \Maryland %
\author{Harald P. Pfeiffer} \Caltech %
\author{Mark A. Scheel} \Caltech %
\author{Luisa T. Buchman} \Caltech %
\author{Lawrence E. Kidder} \Cornell %

\begin{abstract} 
We calibrate the effective-one-body (EOB) model to an accurate
numerical simulation of an equal-mass, non-spinning binary black-hole
coalescence produced by the Caltech-Cornell
collaboration.  Aligning the EOB and numerical waveforms at low
frequency over a time interval of $\sim 1000M$,
and taking into account the uncertainties in the
numerical simulation, we investigate the significance and degeneracy 
of the EOB adjustable parameters during inspiral, plunge and merger, 
and determine the minimum number of EOB adjustable parameters that achieves phase and
amplitude agreements on the order of the numerical error.  We find
that phase and fractional amplitude differences 
between the numerical and EOB values of the dominant gravitational wave mode $h_{22}$ can be reduced to
$0.02$ radians and $2\%$, respectively, until a time $20 M$
before merger, and to $0.04$ radians and $7\%$, respectively,
at a time $20 M$ after merger 
(during ringdown). 
Using LIGO, Enhanced LIGO and Advanced LIGO noise curves,
we find that the overlap between the EOB and the numerical 
$h_{22}$, maximized only over the initial phase and 
time of arrival, is larger than $0.999$ for equal-mass binary black holes
with total mass $30 \mbox{--} 150 M_\odot$.
In addition to the leading gravitational mode $(2,2)$, 
we compare the
dominant subleading modes $(4,4)$ and $(3,2)$ for the inspiral
and find phase 
and amplitude differences on the order of the numerical error. 
We also determine the mass-ratio dependence of one of the EOB adjustable
parameters by calibrating to numerical {\it inspiral} waveforms
for black-hole binaries with mass ratios 2:1 and 3:1. 
The results presented in this paper improve
and extend recent successful attempts aimed at providing gravitational-wave
data analysts the best analytical EOB  model capable of
interpolating accurate numerical simulations.
\end{abstract}

\date{\today \hspace{0.2truecm}}


\maketitle

\section{Introduction}
\label{sec:intro}

The first-generation gravitational-wave
detectors --- the Laser Interferometer Gravitational Wave
Observatory (LIGO)~\cite{Barish:1999,Waldman:2006}, 
GEO~\cite{Hild:2006} and Virgo~\cite{Acernese-etal:2006} --- 
have operated at design sensitivity for a few years, providing new upper
limits for several astrophysical sources. They are now undergoing an
upgrade to Enhanced LIGO and Virgo$^+$; this will improve their
sensitivity by a factor of $\sim 2$. The second-generation
interferometers, Advanced LIGO~\cite{Fritschel2003} and Advanced Virgo,
will start operating in 2013-2015 with an  overall improvement in
sensitivity by a factor of $\sim 10$, thus increasing the event rates
for many astrophysical sources by a factor of one thousand.

One of the most promising sources for these detectors is the inspiral
and merger of compact binary systems of black holes. The search for
gravitational waves (GWs) from coalescing binaries and the extraction
of parameters are based on the matched-filtering
technique~\cite{Finn1992,Finn1993}, which requires a rather accurate
knowledge of the waveform of the incoming signal~\cite{Lindblom2008}.
In particular, the detection and subsequent data analysis of GW
signals are made by using a bank of templates modeling the GWs emitted
by the source.

The effective-one-body (EOB) formalism was
introduced~\cite{Buonanno99,Buonanno00} as a promising approach to
describe analytically the inspiral, merger, and ringdown waveforms
emitted during a binary merger.  Necessary inputs for the EOB approach
include high-order post-Newtonian (PN) results~\cite{Blanchet2006}
for two-body conservative dynamics,
radiation-reaction force, and gravitational waveforms.
For compact bodies, the PN approximation is essentially an expansion in
the characteristic orbital velocity $v/c$ or, equivalently, in the
gravitational potential, $GM/(rc^2)$, with $r$ the typical separation
and $M$ the total binary mass.  The EOB approach, however, does not
use the PN results in their original {\it Taylor-expanded} forms (i.e., as polynomials in
$v/c$), but instead in some {\it resummed}
forms~\cite{Damour98,Buonanno99,Buonanno00,DJS00,Damour01c,Damour03,Damour2007,
  DIN}.  The latter are designed to incorporate some of the expected
non-perturbative features of the exact results.

As it is now possible to produce very accurate numerical simulations
of comparable mass binary black-hole coalescences (see
e.g.~\cite{Bruegmann2006,Husa2007,
Hannam2007,Boyle2007,Boyle2008a,Scheel2008,Vaishnav-Hinderer:2007,Hannam:2009hh}), we can
compare in detail the EOB predictions with numerical results, and when 
necessary, introduce new features into the EOB model in order to improve 
its agreement with the numerical results.
This is an important avenue to LIGO, GEO and Virgo
template construction, as eventually thousands of waveform templates
may be needed to detect the GW signal within the detector noise, and
to extract astrophysical information from the observed waveform. Given
the high computational cost of running the numerical simulations,
template construction is currently an impossible demand for numerical
relativity alone.

This paper builds upon a rather successful recent effort~\cite{Buonanno-Cook-Pretorius:2007,Pan2007,
  Buonanno2007,Damour2007a,DN2007b,DN2008,Boyle2008a} aimed at
producing the best analytical EOB model able to interpolate accurate
numerical simulations. Other approaches based on phenomenological
waveforms have also been
proposed~\cite{Ajith-Babak-Chen-etal:2007,Ajith-Babak-Chen-etal:2007b}.
Here we calibrate the EOB model to the most accurate numerical
simulation to date of an equal-mass, non-spinning binary black-hole merger, 
that has been produced with a pseudospectral code by the Caltech-Cornell
collaboration~\cite{Boyle2007,Scheel2008}. Taking into account the uncertainties 
in the numerical simulation, we investigate the significance and degeneracy 
of the EOB adjustable parameters and determine the minimal 
number of adjustable parameters that 
achieves as good agreement as possible between the numerical and EOB GW's phase and
amplitude. In addition to the leading GW mode $(\ell,m) = (2,2)$, we also 
compare the leading subdominant modes $(4,4)$ and $(3,2)$.  By
reducing the phase difference between the EOB and numerical {\it
  inspiral} waveforms of black-hole binaries with mass ratios $q=m_1$:$m_2$ of
2:1 and 3:1, we explore the dependence of one of the adjustable parameters on
the symmetric mass ratio $\nu =m_1\,m_2/(m_1+m_2)^2$. 

The paper is organized as follows. In Sec.~\ref{sec:EOB}, we briefly
review the EOB dynamics and waveforms. In
Sec.~\ref{sec:EOBcalibration}, we calibrate the EOB model to the
numerical simulation of an equal-mass non-spinning binary black-hole
coalescence and determine the region of the parameter space of the EOB
adjustable parameters that leads to the best agreement with the
numerical results. We also discuss the impact of our results 
on data analysis, and calibrate the EOB model with inspiral 
waveforms from accurate numerical simulations of non-spinning black hole 
binaries with mass ratios 2:1 and 3:1.
Sec.~\ref{sec:conclusions} summarizes our main
conclusions. Finally, the Appendix
compares the
numerical $h_{\ell m}$ extracted with the Regge-Wheeler-Zerilli (RWZ) formalism
with the $h_{\ell m}$ obtained by two time integrals of the Newman-Penrose (NP) 
scalar $\Psi_4^{\ell m}$.

\section{Effective-one-body model}
\label{sec:EOB}

In this section we briefly review the EOB dynamics and waveforms,
focusing mainly on the adjustable parameters. More details can be
found in Refs.~\cite{Buonanno99,DJS00,Buonanno00,Damour03,Buonanno-Cook-Pretorius:2007,
  Buonanno2007,Damour2007a,DN2007b,DN2008,Boyle2008a}. Here we follow
Refs.~\cite{Buonanno2007,Boyle2008a}.

\subsection{Effective-one-body dynamics}
\label{sec:EOBdynamics}

We set $M=m_1+m_2$, $\mu = m_1\,m_2/M = \nu \,M$, and use natural
units $G=c=1$. In absence of spins, the motion is constrained to a
plane. Introducing polar coordinates $(r,\OrbitalPhase)$ and their
conjugate momenta $( p_r,p_\OrbitalPhase)$, the EOB effective metric
takes the form~\cite{Buonanno99}
\begin{equation}
  ds_{\rm eff}^2 =
  -A(r)\,dt^2 + \frac{D(r)}{A(r)}\,dr^2 +
  r^2\,\Big(d\theta^2+\sin^2\theta\,d\OrbitalPhase^2\Big) \,.  
\label{eq:EOBmetric}
\end{equation}
Following Ref.~\cite{Damour2007,Damour:2007cb}, we replace the radial
momentum $p_r$ with $p_{r_*}$, the conjugate momentum to the
EOB {\it tortoise} radial coordinate $r_*$:
\begin{equation}
\frac{dr_*}{dr}=\frac{\sqrt{D(r)}}{A(r)}\,.
\end{equation}
In terms of $p_{r_*}$ the non-spinning EOB Hamiltonian
is~\cite{Buonanno99}
\begin{multline}
\label{himpr}
H^{\rm real}(r,p_{r_*},p_\OrbitalPhase) \equiv \mu\hat{H}^{\rm real} \\
= M\,\sqrt{1 + 2\nu\,\left ( \frac{H^{\rm eff} - \mu}{\mu}\right )} -M\,,
\end{multline}
with the effective Hamiltonian~\cite{Buonanno99,DJS00,Damour:2007cb}
\begin{multline}
  \label{eq:genexp}
  H^{\rm eff}(r,p_{r_*},p_\OrbitalPhase) \equiv \mu\,\widehat{H}^{\rm eff} \\
  = \mu\,\sqrt{p^2_{r_*}+A (r) \left[ 1 + \frac{p_\OrbitalPhase^2}{r^2} 
      + 2(4-3\nu)\,\nu\,\frac{p_{r_*}^4}{r^2} \right]} \,.
\end{multline}
The Taylor-approximants to the coefficients $A(r)$ and $D(r)$ can be
written as~\cite{Buonanno99,DJS00}
\begin{eqnarray}
  \label{coeffA}
  A_{k}(r) &=& \sum_{i=0}^{k+1} \frac{a_i(\nu)}{r^i}\,,\\
  \label{coeffD}
  D_{k}(r) &=& \sum_{i=0}^k \frac{d_i(\nu)}{r^i}\,.
\end{eqnarray}
The functions $A(r)$, $D(r)$, $A_k(r)$ and $D_k(r)$ all depend on 
the symmetric mass ratio $\nu$ through the $\nu$--dependent coefficients 
$a_i(\nu)$ and $d_i(\nu)$.  These coefficients are currently
known through 3PN order (i.e. up to $k=4$) and can be read 
from Eqs.~(47) and (48) in Ref.~\cite{Boyle2008a}. 
Previous investigations~\cite{Damour03,Buonanno2007,Damour2007a,DN2007b,DN2008,
Boyle2008a} have demonstrated that, during the last stages of inspiral
and plunge, the EOB dynamics can be adjusted closer to the numerical
simulations by including in the radial potential $A(r)$ a pseudo 4PN
(p4PN) coefficient 
$a_5(\nu)$.  This coefficient has so far been treated as a
  {\em linear} function in $\nu$, i.e. $a_5(\nu)=\lambda_0\,\nu$, with
  $\lambda_0$ a constant\footnote{Note that $\lambda_0$ 
    was denoted $\lambda$ in Ref.~\cite{Buonanno2007}, and $a_5$ in 
Refs.~\cite{Damour2007a,DN2007b,DN2008,
Boyle2008a}.}.
  In this paper, however, we shall also explore the
  possibility of going beyond this linear dependence, such that
\begin{equation}\label{eq:a5}
a_5(\nu)=\nu\left(\lambda_0+\lambda_1\,\nu\right)\,,
\end{equation}
where $\lambda_0$ and $\lambda_1$ are constants. 
In order to assure the presence of a horizon in the effective metric (\ref{eq:EOBmetric}), a zero needs to
be factored out from $A(r)$. This is obtained by applying a \pade
resummation~\cite{DJS00}. The \pade coefficients for the expansion of
$A(r)$ and $D(r)$ at p4PN order are denoted  
$A_4^1(r)$ and $D_4^0(r)$, and their explicit form can
be read from Eqs.~(54) and (59) in Ref.~\cite{Boyle2008a}.

The EOB Hamilton equations are written in terms of the reduced (i.e., dimensionless) quantities
  $\widehat{H}^{\rm real}$ [defined in Eq.~(\ref{himpr})], $\widehat{t} = t/M$,
and  $\widehat{\OrbitalFreq} = \OrbitalFreq\,M$~\cite{Buonanno00}: 
\begin{eqnarray}
  \frac{dr}{d \widehat{t}} &=& 
  \frac{A(r)}{\sqrt{D(r)}}\frac{\partial \widehat{H}^{\rm real}}
  {\partial p_{r_*}}(r,p_{r_*},p_\OrbitalPhase)\,, 
  \label{eq:eobhamone} \\
  \frac{d \OrbitalPhase}{d \widehat{t}}  &=& 
  \frac{\partial \widehat{H}^{\rm real}}
  {\partial p_\OrbitalPhase}(r,p_{r_*},p_\OrbitalPhase)\,, \\
  \frac{d p_{r_*}}{d \widehat{t}} &=& 
  \frac{A(r)}{\sqrt{D(r)}} \left [- \frac{\partial \widehat{H}^{\rm real}}
  {\partial r}(r,p_{r_*},p_\OrbitalPhase) 
  \label{eq:eobhamthree}
+ \widehat{\cal F}_r(r,p_{r_*},p_{\OrbitalPhase}) \right ], \nonumber \\ && \\
  \frac{d p_\OrbitalPhase}{d \widehat{t}} &=&
  \widehat{\cal F}_\OrbitalPhase(r,p_{r_*},p_{\OrbitalPhase})\,,
  \label{eq:eobhamfour}
\end{eqnarray}
with the definition $\widehat{\OrbitalFreq}\equiv d \OrbitalPhase/d \widehat{t}$.  Furthermore, 
for the $\OrbitalPhase$ component of the radiation-reaction
force we use the non-Keplerian \pade-approximant to the energy
flux~\cite{Damour98,Damour06}
\begin{equation}
    \label{fluxPnK}
\widehat{\cal F}_\OrbitalPhase=\;  
{}^{\rm nK} \widehat{\cal F}_4^4 \equiv  - 
    \frac{v^3_\Omega}{\nu V_\OrbitalPhase^6}\,
    {F}_{4}^4(V_\OrbitalPhase;\nu, v_{\rm pole})\,,
\end{equation}
where $v_\Omega\!\equiv\!\widehat{\OrbitalFreq}^{1/3}$, 
$V_\OrbitalPhase\!\equiv\!\widehat{\OrbitalFreq}\,r_\OrbitalFreq$,
and $r_\OrbitalFreq\!\equiv\!r\,[\psi(r,p_\OrbitalPhase)]^{1/3}$. 
Here $\psi$ is defined by Eqs.~(66)--(68) of Ref.~\cite{Boyle2008a}. 
As the EOB Hamiltonian is a deformation of the Schwarzschild Hamiltonian, the exact Keplerian
relation $\widehat{\OrbitalFreq}^2\,r_\OrbitalFreq^3 = 1$ holds. 
The quantity $F_4^4$ in Eq.~(\ref{fluxPnK}) is given by Eqs.~(39) and (40)
in Ref.~\cite{Boyle2008a}\footnote{Note that here we use the \pade
approximants with factorized logarithms, as originally proposed in
Ref.~\cite{Damour98}, but we set $v_{\rm LSO} =1$, so that the GW energy
flux depends only on the two adjustable parameters $v_{\rm pole}$ and
$A_8$.}  and it uses the Taylor-expanded energy flux (as given by Eq.~(19) in Ref.~\cite{Boyle2008a}) in the form
\begin{eqnarray}
{\cal F}_{8}(\nu) &=&
-\frac{323105549467}{3178375200} + \frac{232597}{4410}\gamma_E 
-\frac{1369}{126}\pi^2 \nonumber\\ &&
+ \frac{39931}{294}\log 2 - \frac{47385}{1568}\log3 + 
\frac{232597}{4410}\log v_\Omega \nonumber\\ &&+ \nu A_8
\,,
\end{eqnarray}
where we combine the known test-mass-limit terms~\cite{Tagoshi94} with
a p4PN adjustable parameter $A_8$~\cite{Boyle2008a}.\footnote{Note that 
in Ref.~\cite{Boyle2008a} the p4PN contribution in the GW energy flux also 
included the term $\nu B_8 \log v_\Omega$. Since we found appreciable 
degeneracy between $A_8$ and $B_8$, we disregard $B_8$, i.e., we set $B_8=0$.}
  
The radial component of the radiation-reaction force $\widehat{\cal
F}_r(r,p_{r_*},p_{\OrbitalPhase})$ in Eq.~(\ref{eq:eobhamthree})  was
neglected in previous studies~\cite{Buonanno-Cook-Pretorius:2007,Buonanno2007,
Damour2007a,DN2007b,DN2008,Boyle2008a}
because Ref.~\cite{Buonanno00} showed that for quasi-circular motion, in some gauges,
it can be set to zero. Furthermore, it was shown in Ref.~\cite{Buonanno00} that 
if the motion remains quasi-circular even during the plunge, $\widehat{\cal
F}_r(r,p_{r_*},p_{\OrbitalPhase})$ does not affect the dynamics
considerably. However, since we are trying to capture effects in the
numerical simulations which go beyond the quasi-circular motion
assumption, we find it interesting to add $\widehat{\cal
F}_r(r,p_{r_*},p_{\OrbitalPhase})$ [see Eq.~(3.18) of Ref.~\cite{Buonanno00} and 
the discussion around it]. 
We set
\begin{equation}
\widehat{\cal F}_r(r,p_{r_*},p_{\OrbitalPhase}) = 
a_{\rm RR}^{{\cal F}_r}(\nu)\,\frac{\dot{r}}{r^2 \Omega}\,
\widehat{\cal F}_\OrbitalPhase(r,p_{r_*},p_{\OrbitalPhase})\,, 
\label{eq:Fluxr}
\end{equation}
where $a_{\rm RR}^{{\cal F}_r}(\nu)$ is an adjustable parameter.

Finally, the tangential force described by Eq.~(\ref{fluxPnK}) applies only
to quasi-circular motion.  This tangential force could also in principle
contain terms describing the departure from quasi-circular motion during
the last stages of inspiral and plunge.  There are several ways to include
such non-quasi-circular (NQC) terms~\cite{Damour03,DN2008,Boyle2008a}; 
here we do so by replacing the
quantity $\widehat{\cal F}_\OrbitalPhase$ on the right-hand side of
Eq.~(\ref{eq:eobhamfour}) [but not the  $\widehat{\cal F}_\OrbitalPhase$
on the right-hand side of Eq.~(\ref{eq:Fluxr})] 
with ${}^{\rm NQC}\widehat{\cal F}_\Phi$, where
\begin{equation}
\label{fluxNQC}
{}^{\rm NQC}\widehat{\cal F}_\Phi \equiv 
\widehat{\cal F}_\Phi\,\left ( 1 + 
a^{{\cal F}_\Phi}_{\rm RR}(\nu)\,\frac{\dot{r}^2}{(r \Omega)^2} 
\right )\,,
\end{equation}
and $a^{{\cal F}_\Phi}_{\rm RR}(\nu)$ is an additional adjustable
parameter.  The form of this NQC correction will be discussed
further in Sec.~\ref{sec:EOBtuning}.
Note that alternative NQC terms have been
proposed in the literature---for example, in Ref.~\cite{Damour03} the
authors used $p_{r}^2/(p_\Phi/r^2)$ while Ref.~\cite{DN2008} employed
$p_{r_*}^2/(r \Omega)^2$. In summary, in the notation of
Ref.~\cite{Boyle2008a}, the EOB model used here is ${}^{\rm
nK}F_4^4/H_{4}$ with adjustable parameters 
$\{a_5(\nu),v_{\rm pole}(\nu),
a_{\rm RR}^{{\cal F}_\Phi}(\nu),a_{\rm RR}^{{\cal F}_r}(\nu),A_8\}$.

\subsection{EOB waveform: Inspiral \& Plunge}
\label{sec:EOBinspiralwaveforms}

Having the inspiral dynamics in hand, we need to compute the gravitational waveform $h_{\ell m}$. 
Reference~\cite{Boyle2007} compared the numerically extracted gravitational waveform $h_{22}$ to the PN 
result with amplitude expressed as a 
{\em Taylor}--expansion~\cite{Kidder:2007rt,BFIS}; even when expanded 
to 3PN order, the amplitude disagreed by about one percent at times 
several hundred $M$ before merger. As previous investigations~\cite{Damour2007a,DN2007b,DN2008} have shown, 
more accurate agreement with the numerical $h_{22}$ amplitude can be
obtained by applying several resummations to the Taylor-expanded
$h_{22}$ amplitude.   These resummations
have recently been improved using results in the quasi-circular test-particle
limit~\cite{DIN}. We follow Ref.~\cite{DIN} and write the EOB modes
$h_{\ell m}$ as
\begin{subequations} 
\begin{eqnarray}
\label{h22}
\widehat{h}_{22}(t) &=&-\frac{8M}{R} \sqrt{\frac{\pi }{5}}\,\nu\,e^{-2 i \OrbitalPhase }\,V_\OrbitalPhase^2\,F_{22}\,,\\
\widehat{h}_{44}(t) &=&-\frac{64 M}{9R}\,\sqrt{\frac{\pi}{7}}\,\nu(1-3\nu)\,e^{-4 i \OrbitalPhase }\,V_\OrbitalPhase^4\,F_{44}\,,\\
\widehat{h}_{32}(t) &=&-\frac{8M}{3R}\,\sqrt{\frac{\pi}{7}}\,\nu(1-3\nu)\,e^{-2 i \OrbitalPhase }\,V_\OrbitalPhase^4\,F_{32}\,,
\label{h32}
\end{eqnarray}
\end{subequations} 
where $R$ is the luminosity distance from the binary, and with
\begin{equation}
F_{lm} = 
\begin{cases}
\hat{H}_{\rm eff}\, T_{\ell m}\, (\rho_{\ell m})^\ell\, 
e^{{\rm i}\delta_{\ell m}} & \text{($\ell + m$ even)}\\
\hat{J}_{\rm eff}\, T_{\ell m}\, (\rho^{J}_{\ell m})^\ell\, 
e^{{\rm i}\delta_{\ell m}}  & \text{($\ell + m$ odd)}
\end{cases}
\end{equation} 
where $\hat{H}_{\rm eff}$ and $\hat{J}_{\rm eff}$ are 
effective sources that in the test-particle, circular-motion limit
contain a pole at the EOB light ring (photon orbit); here
$\hat{H}_{\rm eff}$ is given in Eq.~(\ref{eq:genexp}), and
$\hat{J}_{\rm eff}=p_\OrbitalPhase v_\Omega$ is equal to the orbital angular 
momentum $p_\OrbitalPhase$ normalized to the
circular-orbit Newtonian angular momentum $v_\Omega^{-1}$.
The quantities $T_{\ell m},
\delta_{\ell m}, \rho_{\ell m}, \rho_{\ell m}^J$ can be read from
Eqs.~(19), (20), (23), (25), (C1), (C4) and (C6) in Ref.~\cite{DIN},
respectively.  More specifically, $T_{\ell m}$ is a resummed
version~\cite{Damour2007} of an infinite number of leading logarithms
entering the tail effects; $\delta_{\ell m}$ is a supplementary
phase~\cite{Damour2007} which corrects the phase effects not included
in the complex tail factor; $\rho_{\ell m}$ and $\rho_{\ell m}^J$ are
the resummed expressions of higher-order PN effects as recently
proposed in Ref.~\cite{DIN} in the test-particle circular-orbit
limit. The latter resummation was proposed to cure, among other
effects, the linear growth with $\ell$ of the 1PN corrections in the
Taylor-expanded amplitude.  

Furthermore, motivated by the PN expansion for generic orbits, to include NQC effects in $h_{\ell m}$ we write
%
%
\begin{eqnarray}\label{inspwavenew}
&& {h}_{\ell m}^{\rm insp-plunge} \equiv {}^{\rm NQC}{h}_{\ell m} =
\widehat{h}_{\ell m}\,\left [ 1 + a^{h_{\ell m}}_{1}\,
\frac{\dot{r}^{2}}{(r \Omega)^{2}} \right .\nonumber \\
&& \left . + \dot{r}^2\,\left (a^{h_{\ell m}}_{2}\,\frac{\dot{r}^2}{(r \Omega)^{2}} + 
a^{h_{\ell m}}_{3}\,\frac{M}{r}\,\frac{1}{(r \Omega)^{2}} \right ) \right . \nonumber \\
&& \left . + \dot{r}^4\,a^{h_{\ell m}}_{4}\,\frac{M}{r}\,\frac{1}{(r \Omega)^{2}} \right ]\,.
\end{eqnarray}
As we shall discuss in detail below, for the (2,2) mode, one 
of the four adjustable parameters 
$a^{h_{22}}_i$ in Eq.~(\ref{inspwavenew}) will be fixed by requiring that
the peak of the EOB $h_{22}$ occurs at the same time as the peak of
the EOB orbital frequency~\cite{DN2008} (i.e., at the EOB light-ring);
this requires no matching to a numerical waveform.  Another of the
$a^{h_{22}}_i$ will be fixed by requiring that the peak amplitude of
the EOB and numerical waveforms agree.  The final three $a^{h_{22}}_i$
parameters will be determined by minimizing the overall amplitude
difference with respect to the numerical waveform.  
We note that an alternative NQC factor has been proposed in
Ref.~\cite{DN2008}, notably $1 + a\, p_{r_\star}^2/(\Omega^2\,r^2
+\epsilon)$. We shall compare those different choices below.

\subsection{EOB waveform: Merger \& Ringdown}
\label{sec:EOBmergerRDwaveform}

The merger-ringdown waveform in the EOB approach is built as
follows~\cite{Buonanno00,Damour06,
Buonanno-Cook-Pretorius:2007,Buonanno2007,DN2007b,DN2008}. For each
mode $(\ell,m)$ we write
\begin{equation}
\label{RD}
h_{\ell m}^{\rm merger-RD}(t) = \sum_{n=0}^{N-1} A_{\ell mn}\,
e^{-i\sigma_{\ell mn} (t-t_{\rm match}^{\ell m})},
\end{equation}
where $n$ is the overtone number of the Kerr
quasi-normal mode (QNM), $N$ is the number of overtones included in
our model, and $A_{\ell mn}$ are complex amplitudes to be determined
by a matching procedure described below. 
The quantity $\sigma_{\ell mn} = \omega_{\ell m n} - i
\alpha_{\ell m n}$, where the oscillation frequencies $\omega_{\ell m n}>0$ and the inverse decay-times $\alpha_{\ell m
n}>0$, 
are numbers associated with each QNM. The
complex frequencies are known functions of the final black-hole mass
and spin and can be found in Ref.~\cite{Berti2006a}. The final
black-hole masses and spins can be obtained from several fitting
formulae to numerical results~\cite{Berti2007,Buonanno2007,Damour:2007cb,RezollaBarausse2008}.
Here we use the more accurate final black-hole mass and spin computed in Ref.~\cite{Scheel2008}: 
$M_{\rm BH}/M = 0.95162 \pm 0.00002$, $a/M_{\rm BH} = 0.68646 \pm
0.00004$. While these numbers differ from the predictions of the fitting formulae in
Ref.~\cite{Buonanno2007} by only $0.3\%$, such disagreement 
would be noticeable in our comparison. 
The matching time $t_{\rm match}^{22}(\nu)$ is an adjustable parameter that will be 
chosen to be very close to the EOB light-ring~\cite{Buonanno00} when matching the
mode $h_{22}$. 

The complex amplitudes $A_{\ell mn}$ in Eq.~(\ref{RD}) are determined
by matching the EOB merger-ringdown waveform with the EOB
inspiral-plunge waveform. In order to do this, $N$ independent complex
equations are needed. In
Refs.~\cite{Buonanno00,Damour06,Buonanno-Cook-Pretorius:2007,Schnittman2007,
Buonanno2007}, the $N$ equations were obtained at the matching time by
imposing continuity of the waveform and its time derivatives
\begin{multline}
\frac{d^k}{dt^k}h_{\ell m}^{\rm insp-plunge}(t_{\rm match}^{\ell m})=
\frac{d^k}{dt^k}h_{\ell m}^{\rm merger-RD}(t_{\rm match}^{\ell m})\,,\\ 
(k=0,1,2,\cdots, N-1)\,,
\end{multline}
and we denote this approach {\it point matching}. In
Refs.~\cite{DN2007b,DN2008}, the {\it comb matching} approach was
introduced. In this approach, $N$ equations are obtained at $N$ points
evenly sampled in a small time interval $\Delta t_{\rm match}^{\ell
m}$ centered at $t_{\rm match}^{\ell m}$
\begin{multline}
h_{\ell m}^{\rm insp-plunge}(t_{\rm match}^{\ell m}
+ \frac{2k-N+1}{2N-2}\Delta t_{\rm match}^{\ell m})\\=
h_{\ell m}^{\rm merger-RD}(t_{\rm match}^{\ell m}
+ \frac{2k-N+1}{2N-2}\Delta t_{\rm match}^{\ell m} )\,,\\ 
(k=0,1,2,\cdots, N-1)\,.
\end{multline}
Finally, the full (inspiral-plunge-merger-ringdown) EOB waveform reads
\begin{equation}
\label{eobfullwave}
h_{\ell m} = h_{\ell m}^{\rm insp-plunge}\,
\theta(t_{\rm match}^{\ell m} - t) + 
h_{\ell m}^{\rm merger-RD}\,\theta(t-t_{\rm match}^{\ell m})\,.
\end{equation}
The point matching approach gives better smoothness around the
matching time, but it is not very stable numerically when $N$ is large
and higher order numerical derivatives are needed. As we include 
eight QNMs in our ringdown waveforms, we find that 
the comb matching approach is more stable. To improve 
the smoothness of the comb matching we use here 
a {\it hybrid comb matching}: We choose a time interval $\Delta t_{\rm match}^{\ell
    m}$ ending at $t_{\rm match}^{\ell m}$, we impose the continuity of 
the waveform at $N-4$ points evenly sampled from $t_{\rm match}^{\ell m}-\Delta
  t_{\rm match}^{\ell m}$ to $t_{\rm match}^{\ell m}$, but we also 
  require continuity of the first and second order time derivatives of
  the waveform at $t_{\rm match}^{\ell m}-\Delta t_{\rm match}^{\ell
    m}$ and $t_{\rm match}^{\ell m}$, thus guaranteeing the continuity of $\ddot{h}_{\ell m}$. 
  Furthermore, we fix $t_{\rm match}^{\ell m}$ to be the time when the EOB orbital
  frequency reaches its maximum, and tune $\Delta t_{\rm match}^{22}$
  in the range $2.5M \mbox{--}3.5M$ depending on the EOB dynamics.

It is worth noticing that the lowest frequency 
among the eight QNMs included  
in our merger-ringdown waveform is $M\omega_{227}\sim 0.44$, which is 
  larger than the EOB inspiral-plunge waveform frequency 
$M\omega(t_{\rm match}^{22}) \sim 
0.36$. Therefore, generically
the EOB GW frequency will grow very rapidly from $M\omega\sim 0.36$ to 
$M\omega\sim 0.44$ immediately after the matching time, and this growth
can be much more rapid than what is seen in the numerical simulation. 
We find that we can avoid this rapid growth by carefully
fine-tuning the matching interval $\Delta t_{\rm match}^{22}$, and this
is what we do for the comparisons presented here. 
Quite interestingly, we find that the 
$h_{22}$ matching can be made much less
sensitive to $\Delta t_{\rm match}^{22}$ if we include a pseudo QNM that has a frequency 
$M\omega(t_{\rm match}^{22}) \sim 0.36$ and a decay time comparable to that 
of the highest overtone $\tau_{227} \sim 0.7M$. We refer to this QNM as 
pseudo because its frequency and decay time do not coincide with 
any of the QNMs 
of our final Kerr BH~\cite{Berti2006a,BCKO03}. Although we do not use this 
pseudo QNM in the present 
analysis, we expect that its inclusion can help when matching higher modes 
of equal and 
unequal mass binaries and we shall consider it in the future.

\section{Calibrating the effective-one-body waveforms to numerical relativity 
simulations}
\label{sec:EOBcalibration}

We shall now calibrate the EOB model against a numerical
  simulation of an equal-mass non-spinning binary black hole.  This
  simulation was presented as run ''30c1/N6'' in Scheel et al.~\cite{Scheel2008}, 
  and the inspiral part of the waveform was used in previous comparisons
  with PN models~\cite{Boyle2007,Boyle2008a}.  In addition to the
  NP scalars $\Psi_4^{\ell m}$ extracted from this
  simulation, we will be using gravitational waveforms $h_{\ell m}$
  extracted with the RWZ formalism~\cite{ReggeWheeler1957,
  Zerilli1970b,Sarbach2001,Rinne2008b}.  
  The Appendix discusses details of the numerical implementation used to obtain
    $h_{\ell m}$ from the RWZ scalars, and presents a comprehensive comparison 
    of the numerical $\Psi_4^{\ell m}$ and RWZ $h_{\ell m}$ waveforms.
  Consistency between the two wave-extraction schemes is
  good, with phase differences less than $0.02$ radians for the
  (2,2)--mode until about a time $20 M$ after the peak of $|h_{22}|$.

  Because we have more experience with the NP scalars 
  during the inspiral, and
  because $\Psi_4^{22}$ appears to behave better than RWZ $h_{22}$
  during ringdown (see Fig.~\ref{fig:WaveformRe22} in the Appendix),
  we prefer to use the numerical $\Psi_4$ data. 
  Therefore, during the
  inspiral phase, we will calibrate the EOB adjustable parameters by
  comparing the second time derivative of EOB $h_{22}$ against the
  numerical $\Psi_4^{22}$. During the plunge-merger phase, when the 
  time derivatives of the waveform vary most rapidly, 
  it is more difficult to calibrate 
  the EOB $\ddot{h}_{22}$ since the resummation techniques in the 
  EOB model were aimed at providing us with the best $h_{22}$.
  Therefore, around time of merger, 
  we shall calibrate the EOB $h_{22}$ to the RWZ
  $h_{22}$. Note also that data analysis is based on $h_{\ell m}$, 
  further motivating our choice to build
  the best EOB model for $h_{\ell m}$. Nevertheless, after
  calibration, in Sec.~\ref{sec:comparinghlms}, we show comparisons of
  the EOB waveforms with both the numerical RWZ $h_{22}$ and
  $\Psi_4^{22}$.
  The ringdown part of the numerical waveform is not used in 
  the calibration of the EOB parameters;
  the QNMs are determined solely from the mass and spin of the final hole.

\subsection{Waveform alignment and uncertainties in numerical waveforms}
\label{sec:NRerrors}

As previous investigations~\cite{Buonanno-Cook-Pretorius:2007,Baker2006d,Baker2006e,
Boyle2007,DN2008} have shown, the phase error between two waveforms depends
crucially on the procedure used to align them in time and phase.  For the {\em inspiral phase},
we shall adopt here the alignment procedure introduced in
Ref.~\cite{Boyle2008a} (see also Ref.~\cite{Ajith-Babak-Chen-etal:2007b}) that consists of minimizing
the quantity
\begin{equation}\label{waveshifts}
\Xi(\Delta t,\Delta\phi)=\int_{t_1}^{t_2}\left[\phi_1(t)-
\phi_2(t-\Delta t)-\Delta\phi\right]^2\,dt\,,
\end{equation}
over a time shift $\Delta t$ and a phase shift $\Delta\phi$, where
$\phi_1(t)$ and $\phi_2(t)$ are the phases of the two waveforms. This
alignment procedure has the advantage of averaging over the numerical
noise and residual eccentricity when aligning numerical and EOB
waveforms. The range of integration $(t_1, t_2)$ is chosen to be as
early as possible, where we expect the PN-based EOB waveform to be
most valid, but late enough so that it is not contaminated by the junk radiation
present in the numerical initial data. Moreover, the range of
integration should be large enough for the integral to average over
noise and residual eccentricity. Here we fix $t_1=1040M$ and
$t_2=2260M$ (measured from the start of the numerical waveform), so
that we include three full cycles of phase oscillations due to
eccentricity.

\begin{figure}
  \includegraphics[width=\linewidth]{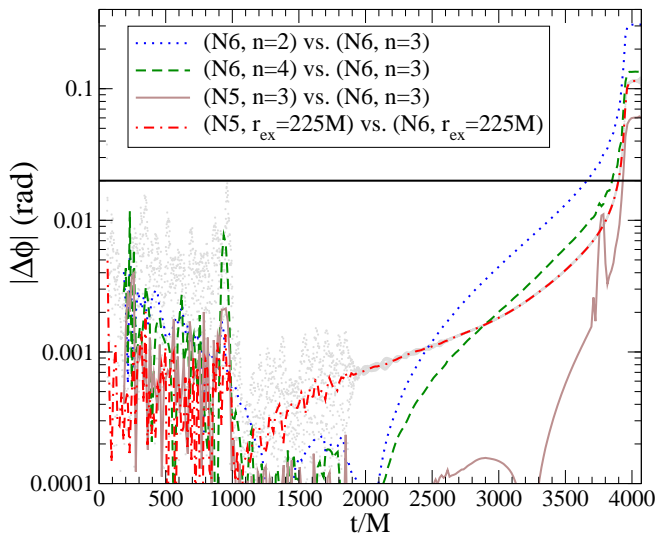}
  \caption{\label{fig:DeltaPhaseNumPsi4} {\bf Numerical error
        estimates.}  Phase difference between numerical $\Psi_4^{22}$
      waveforms, when aligned using the same procedure as employed for
      the EOB-NR alignment [see Eq.~(\ref{waveshifts})].  ``N6'' and
      ``N5'' denote the highest- and next-to-highest numerical
      resolution, $n$ denotes the order of extrapolation to infinite
      extraction radius, and ''$r=225M$'' denotes waves extracted at
      finite radius $r=225M$.  The data are smoothed
      with a rectangular window of
        width $10M$; the light grey dots
          represent the unsmoothed data for the N5-N6 comparison at 
 $r_{\rm ex}=225M$.
      }
  \end{figure}
Using this alignment procedure, we estimate the errors on the
numerical $\Psi_4^{22}$ by comparing $\Psi_4^{22}$ computed at
different numerical resolutions and/or using different extrapolation
procedures. In particular, Fig.~\ref{fig:DeltaPhaseNumPsi4} summarizes
the phase errors for a set of numerical $\Psi_4^{22}$ computed in
Ref.~\cite{Scheel2008}. The numerical waveform labeled ``N6,
n=3'' (identical to the run ``30c1/N6, n=3'' from~\cite{Scheel2008}) 
is the reference numerical waveform used throughout this paper
unless otherwise noted.  This waveform is the most accurate waveform
from Ref.~\cite{Scheel2008}, extracted at various radii and then
extrapolated to infinity.  The waveforms with different values of $n$
vary the order of the extrapolation and are used to quantify the
uncertainty in the phase due to extrapolation, while those labeled by
N5 (as opposed to N6) are from a simulation with a lower numerical resolution and are used to
quantify the uncertainty due to numerical truncation errors.  
Figure~\ref{fig:DeltaPhaseNumPsi4} also includes a comparison 
between waveforms extracted at finite coordinate radius $r_{\rm ex}=225M$. 

Extrapolation with $n=2$ leads to systematic errors in the
  extrapolated waveform (see, Fig.~10 of Ref.~\cite{Boyle2007}), which in
  turn results in a systematic error in $\Delta t$.  Therefore, the
  blue dashed line in Fig.~\ref{fig:DeltaPhaseNumPsi4} represents a
  possibly overly conservative error estimate. The feature of the
  solid brown curve around $t\approx 3700M$ is due to an issue with
  data processing of the lower resolution 'N5' run.

The primary use of Fig.~\ref{fig:DeltaPhaseNumPsi4} is to
  assess numerical errors relevant for the calibration of the EOB {\em
    inspiral} phase.  By construction of the alignment procedure, this
  figure shows the numerical errors for waveforms that are aligned in
  the interval $[t_1, t_2]$, several orbits before merger.
  Calibrating the EOB inspiral phase in this manner is appropriate,
  because it ensures that early in the inspiral, the EOB-model and
  the numerical simulation agree well, i.e. that we expect little
  de-phasing at lower frequencies.  This is important for waveform
  templates of low mass binaries, where the early inspiral waveform
  lies in LIGO's sensitive frequency band.

Figure~\ref{fig:DeltaPhaseNumPsi4} shows that the numerical
  $\Psi_4^{22}$ waveforms are accurate to a few hundredths of a radian until very
  close to merger, when compared with our alignment procedure.  Furthermore,
Fig.~\ref{fig:RWZ-Psi4-Comparison22} in the Appendix demonstrates that 
NP and RWZ waveforms differ
  by only $0.02$ radians through inspiral and merger. 
Therefore, we shall adopt a deviation of $0.02$ radians between EOB- and NR
inspiral-waveforms as our goal for the EOB inspiral calibration.
The horizontal line in Fig.~\ref{fig:DeltaPhaseNumPsi4}
 indicates this phase 
difference of $0.02$ radians and it will be our requirement when
calibrating the EOB values of $\Psi_4^{22}$. The numerical phase
errors exceed $0.02$ radians at times $t=3660M$, $3850M$, 
$3900M$ and $3933M$, respectively, and so our goal will be for EOB to 
agree to $0.02$ radians at least up to $t\approx 3900M$. 
The choice of $0.02$ radians is motivated by the goal of
bringing the disagreement between the EOB and numerical phases {\it at
least} to the level of the numerical error (see
Fig.~\ref{fig:RWZ-Psi4-Comparison22}).

\begin{table}
  \begin{tabular}{|c|c|}
    \hline EOB-dynamics & EOB-waveform\\
     adjustable parameters & adjustable parameters\\\hline
    $a_5(\nu)$ & $t_{\rm match}^{\ell m}(\nu)$\\
    $v_{\rm pole}(\nu)$ & $\Delta t^{\ell m}_{\rm match}(\nu)$\\
    $a_{\rm RR}^{{\cal F}_r}(\nu)$ or $a_{\rm RR}^{{\cal F}_\Phi}(\nu)$ & 
    $a^{h_{\ell m}}_i(\nu) \; i = 1, ... 4$\\
    $A_8$ & \\\hline
    \end{tabular}
    \caption{\label{tableI} Summary of {\it all possible} adjustable parameters of the EOB model 
      considered in this paper. As we shall discuss in the main text, we will not need all of
      these parameters. In particular, we find that for the black-hole binary simulations investigated 
      here, the choices $a_{\rm RR}^{{\cal F}_r}(\nu)=0=a_{\rm RR}^{{\cal F}_\Phi}(\nu), A_8 =0$, 
      $t_{\rm match}^{\ell m}(\nu)$ at the peak of the EOB orbital frequency,
      allow the numerical and EOB values of the GW phase and amplitude to agree within 
      numerical error. Furthermore, we find that for an equal-mass black-hole binary 
      coalescence it is sufficient to set $a_5(\nu) = \nu \lambda_0$ [see Eq.~(\ref{eq:a5}) 
      with $\lambda_1 =0$ ] and calibrate $\lambda_0$, $v_{\rm pole}(1/4)$, 
      $\Delta t^{22 }_{\rm match}(1/4)$ and $a^{h_{22}}_i(1/4)$. For an equal-mass 
      black-hole binary coalescence it is even possible to calibrate {\it only one} EOB-dynamics 
      adjustable parameter, $\lambda_0$ [see Eq.~(\ref{eq:a5})] and let $v_{\rm pole} \rightarrow \infty$. 
      Finally, for an unequal-mass binary inspiral it is sufficient either to set $\lambda_1=0$, use the value of 
      $\lambda_0$ from the equal-mass binary case, and calibrate $v_{\rm pole}(\nu)$; 
      or alternatively to let $v_{\rm pole} \rightarrow \infty$ and calibrate both $\lambda_0$ and $\lambda_1$ 
      in $a_5(\nu)$ [see Eq.~(\ref{eq:a5})].}
\end{table}

\subsection{Tuning the adjustable parameters of the equal-mass effective-one-body dynamics}
\label{sec:EOBtuning}

We divide the adjustable parameters into two groups and tune them
separately in two steps. The first group of EOB-{\it dynamics}
parameters includes $\{a_5(\nu),v_{\rm pole}(\nu), a_{\rm RR}^{{\cal
F}_\Phi}(\nu),a_{\rm RR}^{{\cal F}_r}(\nu),A_8\}$.  These parameters
determine the inspiral and plunge dynamics of the EOB model and affect
the merger-ringdown waveform only indirectly through the waveform's
phase and frequency around the matching point. [We note 
that the inspiral phase is independent of the parameters $a_i^{h_{\ell m}}$, 
see Eq.~(\ref{inspwavenew}).] These parameters are calibrated
to the numerical NP $\Psi_4^{22}$. The second group of
EOB-{\it waveform} parameters includes $\{a^{h_{\ell m}}_i$, $t_{\rm
match}^{\ell m}$, and $\Delta t_{\rm match}^{\ell m}\}$, and affect only
the plunge-merger-ringdown but not the inspiral EOB waveform. 
These parameters are calibrated to the numerical RWZ $h_{22}$.
All the possible adjustable parameters of the EOB model 
employed in this paper are summarized in Table~\ref{tableI}. 
In the first step of
our calibration procedure, we
reduce the phase difference before merger by tuning the EOB-dynamics
parameters. In the second step, we use these fixed values of the
EOB-dynamics parameters, and tune the EOB-waveform parameters.

Among the EOB-dynamics parameters, 
$a_5(\nu)$ and $v_{\rm pole}(\nu)$ are the
most important as they affect the entire quasi-circular evolution of
the inspiral.  The two radiation-reaction parameters $a^{{\cal
F}_\Phi}_{\rm RR}$ and $a^{{\cal F}_r}_{\rm RR}$ are introduced to
adjust the dynamics of late inspiral when we expect that the
quasi-circular assumption is no longer valid. The p4PN parameters in
the energy flux, $A_8$, also influences the entire evolution, but we
find that $A_8$ is strongly degenerate with $a_5(1/4)$ throughout the
inspiral until a time $\sim 100 M$ before merger. 
Based on these considerations, we shall tune $a_5(1/4)$ and $v_{\rm
pole}$ first and consider $a^{{\cal F}_\Phi}_{\rm RR}(1/4)$, $a^{{\cal
F}_r}_{\rm RR}(1/4)$ and $A_8$ only when exploring how to further improve
the late evolution.

Therefore, in our first step, we set  $a^{{\cal F}_\Phi}_{\rm RR}(1/4)=a^{{\cal
F}_r}_{\rm RR}(1/4)=A_8=0$  and vary $a_5(1/4)$ and $v_{\rm pole}(1/4)$. Applying the alignment
procedure presented at the beginning of Sec.~\ref{sec:NRerrors}, 
we shift each EOB $\Psi_4^{22}$ in time and phase to agree with the reference numerical
waveform at low frequency, and determine the time when the phase
difference between the numerical and EOB $\Psi_4^{22}$ waveforms
becomes larger than $0.02$ radians.  We denote this reference time as
$t_{\rm ref}$.  

\begin{figure}
  \includegraphics[width=0.9\linewidth]{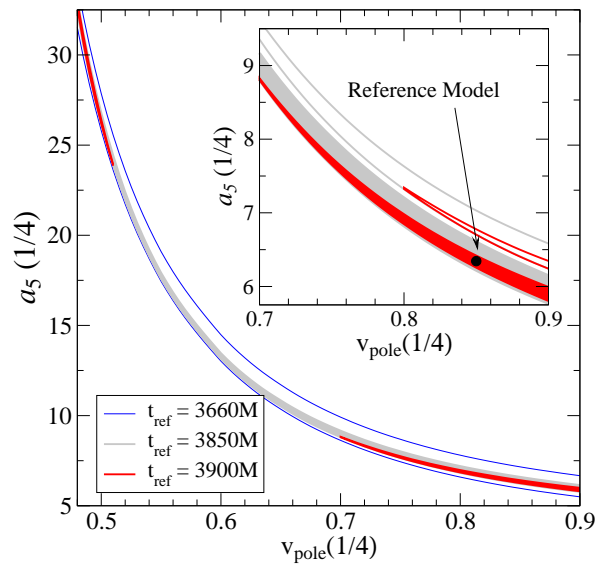}
  \caption{\label{fig:TrefContourPlot4} In the parameter space of the
    EOB-dynamics adjustable parameters $a_5(1/4)$ and $v_{\rm
      pole}(1/4)$ we show the contours of the time $t_{\rm ref}$ at
    which the phase difference between the numerical ``30c1/N6, n=3''
    and EOB $\Psi_4^{22}$ becomes larger than $0.02$ radians.  Note
    that the innermost red contours cover two disjoint regions. The
    inset shows the effect of numerical uncertainty: The filled
    contours are the $t_{\rm ref}=3850M$ and $3900M$ contours from the
    main panel.  The open contours 
    are identical, except computed using the ``30c1/N6, n=2''
    numerical $\Psi_4^{22}$. The reference model is
        shown as a black dot.}
\end{figure}

Figure~\ref{fig:TrefContourPlot4} is a contour plot of the time $t_{\rm
ref}$ in the $a_5(1/4) \mbox{--}v_{\rm pole}(1/4)$ parameter space. For all
points inside the largest contours (blue curves),
the associated EOB $\Psi_4^{22}$ phase evolutions agree with the
numerical ones up to $t=3660M$, which is the earliest reference time 
considered in Sec.~\ref{sec:NRerrors}. In order to get EOB models that have
phase differences less than $0.02$ radians until $t=3900M$, $a_5(1/4)$ and $v_{\rm pole}(1/4)$ have to be inside the
innermost two separate thin contours (red curves).   
One might view these contours as encompassing all values of
$a_5(1/4)$ and $v_{\rm pole}(1/4)$ that are consistent with the numerical inspiral
waveform, given the fixed choices of the various other EOB parameters.
There are $a_5(1/4)$ and $v_{\rm pole}(1/4)$ values that make the EOB phase differences less than $0.03$ 
radians until $t=3933M$, but not less than $0.02$ radians until $t = 3933M$ 
(the latest reference time). We find that phase errors of the EOB $\Psi_4^{22}$
corresponding to the upper left contours in Fig.~\ref{fig:TrefContourPlot4} 
grow rapidly after $t=3900M$, whereas phase errors of EOB $\Psi_4^{22}$ corresponding to the lower right
contours grow only mildly until
around $t=3940M$. For this reason, we shall restrict the
tuning of the other adjustable parameters 
to the lower right region of Fig.~\ref{fig:TrefContourPlot4} 
inside the innermost contour. As a reference set, we choose
$a_5(1/4) = 6.344$  and $v_{\rm pole}(1/4)=0.85$.\footnote{We note that in
Ref.~\cite{Buonanno2007}, the authors suggested as best value $a_5(1/4) = 15$. 
However, the EOB model used in Ref.~\cite{Buonanno2007} differs
from the one employed in this paper, the main difference being the GW
energy flux. More importantly, the procedure used in
Ref.~\cite{Buonanno2007} to calibrate $a_5(1/4)$ was different. It was
based on maximized overlaps with white noise. The best value for $a_5(1/4)$
was obtained by requiring large overlaps, say $\geq\,0.0975$, for
several mass ratios and $(\ell,m)$ modes (see Fig. 2 in
Ref.~\cite{Buonanno2007}). Finally, the accuracy of the numerical
waveforms employed in this paper differ from the ones in
Ref.~\cite{Buonanno2007}.}  We note that the latter value is rather
different from the value obtained in Ref.~\cite{DN2008} when $a_5(1/4) = 6.25$ 
is used.  This is due to differences between the EOB models ---
for example Ref.~\cite{DN2008} employs the \pade-resummed GW energy
flux with constant logarithms, whereas we use the \pade-resummed GW energy
flux with factorized logarithms. 

Quite interestingly, looking more closely at the 
red lines in the right corner of Fig.~\ref{fig:TrefContourPlot4}, as $v_{\rm pole}$ 
increases, we find another possible reference set $a_5(1/4)= 4.19$ and 
$v_{\rm pole} \rightarrow \infty$. With this choice, the pole in the \pade flux 
of Eq.~(\ref{fluxPnK}) disappears.

In order to understand whether further tunings of
radiation-reaction effects by adjusting the parameters $(v_{\rm
pole}, a_{\rm RR}^{{\cal F}_\Phi}, a^{{\cal F}_r}_{\rm RR},A_8)$ can
modify the phasing during plunge, we compute how sensitive the phasing
is to radiation-reaction effects once the binary has passed the last
stable orbit (LSO) defined as $(\partial H^{\rm eff}/\partial r)_{\rm LSO} = 0 = 
(\partial^2 H^{\rm eff}/\partial r^2)_{\rm LSO}$.  
Reference~\cite{BDmg9} pointed out that the phasing during the plunge is not affected much by radiation reaction,
but driven mostly by the conservative dynamics. We want to quantify 
the latter statement more fully.

\begin{figure}
  \includegraphics[width=\linewidth]{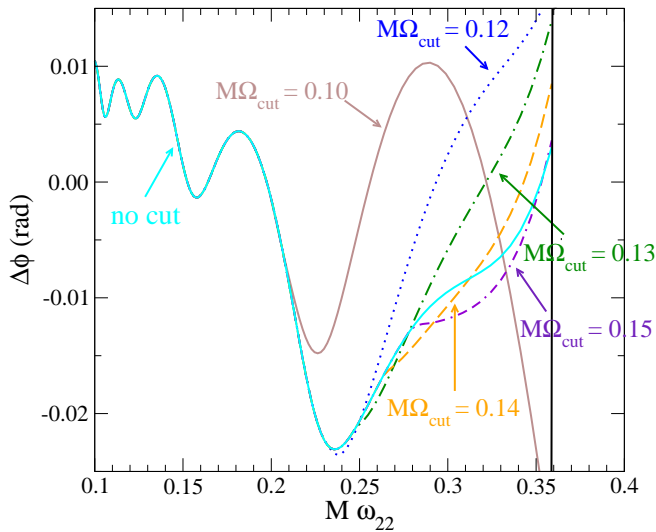}
  \caption{\label{fig:FigFluxOff} For the case $a_5(1/4)=6.344$ and
$v_{\rm pole}(1/4)=0.85$ ($A_8=0$, $a_{\rm RR}^{{\cal F}_\Phi}=0$ and
$a^{{\cal F}_r}_{\rm RR}=0$), we show the phase difference between the
numerical and EOB mode $h_{22}$ versus the numerical GW frequency $M
\omega_{22}$ for EOB models in which the GW energy flux is shut down
at several EOB orbital frequencies. The vertical line marks the
maximum EOB orbital frequency.}
\end{figure}

In order to do this, we need to define when the plunge starts.  
In the absence of radiation reaction, the plunge
starts beyond the LSO where $r=r_{\rm LSO}$, $\omega=\omega_{\rm
LSO}$ and $p_\Phi=p_\Phi^{\rm LSO}$. But in the presence of radiation reaction,
Ref.~\cite{Buonanno00} observed that there is not a unique
$t_{\rm LSO}$ at which the conditions $r=r_{\rm LSO}$,
$\omega=\omega_{\rm LSO}$ and $p_\Phi=p_\Phi^{\rm LSO}$ are
satisfied. In fact, the above conditions may happen at different times
(see Fig. 12 in Ref.~\cite{Buonanno00}). Indeed,  for the case $a_5(1/4)=6.344$ and
$v_{\rm pole}(1/4)=0.85$, we find that with radiation
reaction, $r(t^r_{\rm LSO})=r_{\rm LSO}$,
$\omega(t^\omega_{\rm LSO})=\omega_{\rm LSO}$ and
$p_\Phi(t^{p_\Phi}_{\rm LSO}) = p_\Phi^{\rm LSO}$ where $t^r_{\rm LSO}
=3914.50M$, $t^\omega_{\rm LSO} =3919.83M$ and $t^{p\Phi}_{\rm LSO}
=3885.53M$, and where the orbital frequencies, 
corresponding to the three different $t_{\rm LSO}$ values are
$M \Omega = 0.975, 0.106$, and $0.074$,
respectively.  Following Ref.~\cite{Buonanno00}, we will say that the
plunge starts during the time interval spanned by the values of
$t_{\rm LSO}$ which in this case is $t_{\rm LSO}\sim 34M$ before merger.

In Fig.~\ref{fig:FigFluxOff}, we show the phase difference between the
numerical and EOB $h_{22}$ as a function of the numerical GW frequency
$M\omega_{22} $ for EOB models in which the GW energy flux 
is suddenly shut down at several EOB orbital
frequencies. The cyan curve in Fig.~\ref{fig:FigFluxOff} is obtained 
when the GW energy flux is not shut down. Note that in this case 
the phase difference increases fast close to the EOB matching point, 
which is  marked by the vertical line in Fig.~\ref{fig:FigFluxOff}.
The phase difference can change considerably its shape (including the 
sign of the slope close to the EOB matching point) when the energy 
flux is shut down before $M \Omega = 0.12 \mbox{--}
0.13$, but it does not change much, especially the fast increase close 
to the matching point, when the energy flux is shut down after 
$M \Omega = 0.12 \mbox{--} 0.13$, immediately after the LSO defined 
by the condition
$\omega(t^\omega_{\rm LSO})=\omega_{\rm LSO}$ above.  

This study suggests that it is difficult to modify the behaviour of the EOB
phasing during plunge by tuning only the adjustable parameters
entering the radiation-reaction terms or the GW energy flux, $a_{\rm
RR}^{{\cal F}_r}(\nu)$, $a_{\rm RR}^{{\cal F}_\Phi}(\nu)$ and $A_8$,
$v_{\rm pole}(\nu)$.  The behaviour of the EOB phasing during plunge
is more sensitive to adjustable parameters in the EOB conservative
dynamics, e.g., $a_5(\nu)$ at 4PN order or $a_6(\nu)$ at 5PN order,
etc. However, the parameters $a_i(\nu)$ also affect the phasing during
the very long inspiral, and a careful tuning is needed to reach 
excellent agreement both during inspiral and plunge.

\begin{figure}
  \includegraphics[width=0.9\linewidth]{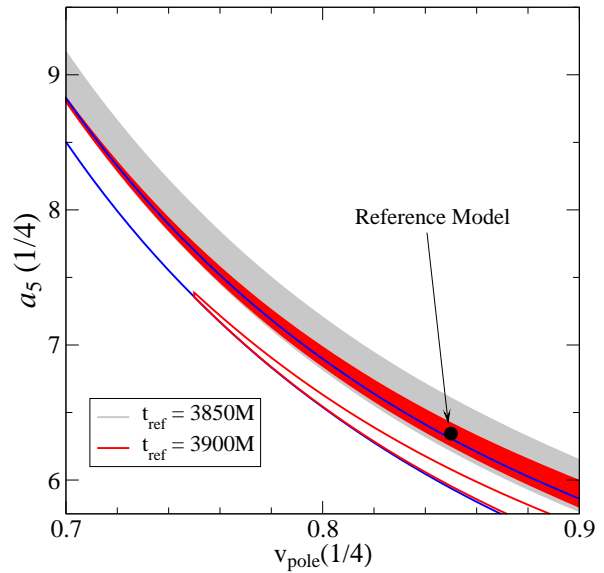}
  \caption{\label{fig:TrefContourPlot5}
Effect of $a^{{\cal F}_r}_{\rm RR}$ on contours of acceptable 
EOB parameters.  The solid contours are the $t_{\rm ref}=3850M$ and $3900M$ contours from Fig.~\ref{fig:TrefContourPlot4}.  
The open contours shifted to the lower-right} are the same, but computed with $a^{{\cal F}_r}_{\rm RR}=0.5$ instead
of $a_{\rm RR}^{{\cal F}_\Phi}=0$. The reference model is shown as a
black dot.
\end{figure}

Nevertheless, it is possible to modify the behaviour of the EOB
phasing during the late inspiral by tuning $A_8$, $a_{\rm RR}^{{\cal
F}_r}(1/4)$ and $a_{\rm RR}^{{\cal F}_\Phi}(1/4)$ {\it together with}
$a_5(1/4)$ and $v_{\rm pole}(1/4)$. As an example of this, we redo the
contour plot shown in Fig.~\ref{fig:TrefContourPlot4}, 
but with $a_{\rm RR}^{{\cal F}_r}(1/4) =0.5$
instead of zero.  The result is shown as dashed curves in Fig.~\ref{fig:TrefContourPlot5}.
We still find EOB models that have phase differences less than $0.02$ radians until
$t=3900M$.  In particular, with the reference value $v_{\rm
pole}(1/4)=0.85$ and choosing $a_5(1/4)= 6.013$, we find that the behaviour
of the EOB phasing is substantially modified only for the last $40M$
of evolution before merger. In this case, the change in phase
difference is in the range of $0.01 \mbox{--} 0.1$ radians, and the
slope of phase difference at the matching point can change
sign. Similar results are obtained when repeating this analysis with $a_{\rm RR}^{{\cal
F}_\Phi}(1/4)$ or $A_8$ different from zero. We also observe that the
effect on the dynamics of the adjustable parameter $a_{\rm RR}^{{\cal
F}_\Phi}(1/4)$ is almost equivalent to the effect of the adjustable
parameter $a_{\rm RR}^{{\cal F}_r}(1/4)$, except for a minus sign and
a different scaling. So it is not necessary to consider both 
of these radiation-reaction adjustable parameters.

Although time consuming, in principle it is possible to perform a
comprehensive search over the complete set of the EOB-dynamics
parameters $a_5(\nu)$, $v_{\rm pole}(\nu)$, $A_8$, $a_{\rm RR}^{{\cal
F}_r}(\nu)$ or $a_{\rm RR}^{{\cal F}_\Phi}(\nu)$.  However, at this
point there is no need to further improve the EOB evolution close to
merger, and achieve better agreement with the equal-mass, non-spinning
numerical data, since the agreement is already at the level of the
numerical error. Thus, in the following, we shall use the values of
$a_5(1/4)$ and $v_{\rm pole}(1/4)$ based on Fig.~\ref{fig:TrefContourPlot4},
obtained by setting to zero all the other EOB-dynamics adjustable
parameters in Table~\ref{tableI}.  We will leave a comprehensive study of the other
EOB-dynamics adjustable parameters to future work when highly accurate
numerical merger waveforms of unequal-mass black-hole binaries become
available.

\begin{figure}
\includegraphics[width=0.95\linewidth]{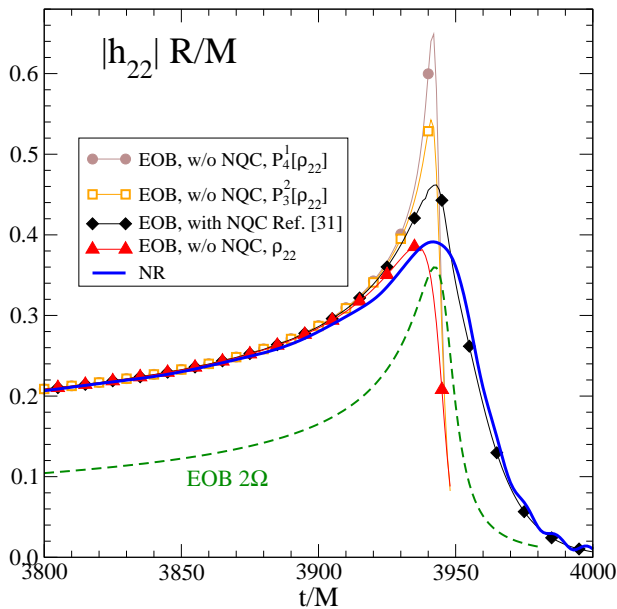}
\caption{\label{fig:h22AmpComparison} We compare the numerical and EOB
$h_{22}$ amplitudes when the EOB model with reference values $a_5(1/4)=6.344$ 
and $v_{\rm pole}(1/4)=0.85$ are used. We show the EOB amplitudes without the NQC
corrections and the EOB amplitude with the NQC terms suggested in
Ref.~\cite{DN2008}, where the NQC parameters take the values $a=0.75$ and $\epsilon=0.09$. 
When the NQC corrections are not included, we show
the EOB amplitude of Eq.~(\ref{h22}) which uses the resummation
procedure of Ref.~\cite{DIN}, and also the EOB amplitudes of
Eq.~(\ref{h22}) when the \pade-resummations $P_4^1$ and $P_3^2$ of
$\rho_{22}$ suggested in Ref.~\cite{DIN} are applied. Note that in
this plot, the EOB amplitudes do not contain the merger-ringdown
contribution.} 
\end{figure}

We shall now discuss the EOB model with reference values $a_5(1/4)=6.344$ 
and $v_{\rm pole}(1/4)=0.85$, and tune the EOB-waveform adjustable parameters. 
We shall comment at the end of this section on the results when the other reference values $a_5(1/4)=4.19$ and 
$v_{\rm pole}(1/4) \rightarrow \infty$ are used.
In Fig.~\ref{fig:h22AmpComparison} we compare the  numerical and 
EOB $h_{22}$ amplitudes with and without including NQC terms.  
The agreement of the numerical amplitude with the EOB 
amplitude of Eq.~(\ref{h22}) without NQC terms, 
which uses the resummation procedure of
Ref.~\cite{DIN}, is rather remarkable. The relative difference at the
peak is only $\sim 1.5\%$, and the EOB peak amplitude occurs only
$\sim 6M$ before the numerical peak amplitude. We notice that this excellent
agreement is due to the presence in $\rho_{22}$ of test-particle
corrections through 5PN order. Were the test-particle corrections
through 4PN or 5PN orders not included, the disagreement at the peak
would become $4.9\%$ and $11.3\%$, respectively.\footnote{In Ref.~\cite{DIN} 
(see Fig. 10 therein and discussion
around it) the authors pointed out that the difference between
$h_{22}$ amplitudes computed with the test-particle corrections
through 3PN, 4PN or 5PN orders, differ {\it only} by a few
percent. However, this statement was obtained for circular orbits
until the LSO frequency $M \Omega =0.097$. Our
Fig.~\ref{fig:h22AmpComparison} extends 
beyond that frequency (the latter corresponds to $t=3914M$ in the figure).} 

Figure~\ref{fig:h22AmpComparison} also shows the EOB amplitudes 
of Eq.~(\ref{h22}) when the \pade-resummations $P_4^1$ and $P_3^2$ of 
$\rho_{22}$ suggested in Ref.~\cite{DIN} are applied. In these cases, the 
EOB peak amplitude almost coincides in time with the numerical peak amplitude, 
but the relative difference in the value of the peak amplitude 
is rather large. However, those large differences may be resolved 
if the resummed version of the GW energy flux~\cite{DIN} consistent with the 
resummed $h_{\ell m}$ were used. 
Figure~\ref{fig:h22AmpComparison} also contains the EOB $h_{22}$
amplitude with NQC terms as suggested in Refs.~\cite{DN2007b,DN2008}
[see Eq. (12) in Ref.~\cite{DN2008}].  The relative
difference with the numerical amplitude is $\sim 20\%$ at the peak.
It is rather interesting to observe, as pointed out in
Ref.~\cite{DN2008}, that by aligning the numerical and EOB waveforms
at low frequency, we find that the peak of the numerical $h_{22}$
coincides with the peak of the EOB orbital frequency.
Here, to improve the amplitude agreement during plunge and merger, we
include the NQC corrections of Eq.~(\ref{inspwavenew}). 
We fix two of the adjustable parameters, $a^{h_{22}}_1$ and $a^{h_{22}}_2$,
by requiring that a local extremum of the EOB $h_{22}$
amplitude occurs at the same time as the peak of the EOB orbital
frequency (i.e., the EOB light-ring), and that the EOB amplitude at
the peak coincides with the numerical amplitude at the peak. In fact,
we expect that in the near future, the peak of the numerical $h_{22}$ 
will be able to be predicted by numerical relativity with high accuracy for
several mass ratios. Thus, the peak can be fit with a polynomial in $\nu$.
(Preliminary studies which use results from Ref.~\cite{Buonanno2007}
confirm this expectation.) The other two adjustable parameters,
i.e., $a^{h_{22}}_3$ and $a^{h_{22}}_4$,
are calibrated to the numerical results to further reduce the
disagreement.  Specifically, we do a two-parameter
least-square-fit of the ratio of the numerical RWZ and EOB $h_{\ell m}$ on 
Eq.~(\ref{inspwavenew}) in which $a^{h_{22}}_1$ and $a^{h_{2}}_2$ are fixed as functions of $a^{h_{22}}_3$ and $a^{h_{2}}_4$ by the requirements described above.  We notice that the strategy of improving the amplitude
agreement followed in this paper might change in the future, when 
accurate numerical unequal-mass black-hole binary inspiral-merger-ringdown waveforms
become available. A smaller number of adjustable parameters might suffice 
if more requirements on the EOB model itself can be imposed
or if a different matching procedure, such as the one
suggested in Ref.~\cite{Baker2008a}, is employed.

\subsection{Comparing the gravitational-wave modes \boldmath$h_{\ell m}$ of 
equal-mass \emph{coalescing} black-hole binaries}
\label{sec:comparinghlms}

In this section, we focus on the model whose EOB-dynamics and
EOB-waveform adjustable parameters were calibrated to numerical RWZ
$h_{22}$ and NP $\Psi_4^{22}$ in Sec.~\ref{sec:EOBtuning}.  Using this
EOB model, we generate the GW modes $h_{22}$, $h_{32}$ and
$h_{44}$ and compare them to the corresponding
numerical modes.  We choose these three modes because they are the
most dominant ones for an equal-mass, non-spinning black-hole binary. 
\begin{figure}
  \includegraphics[width=\linewidth]{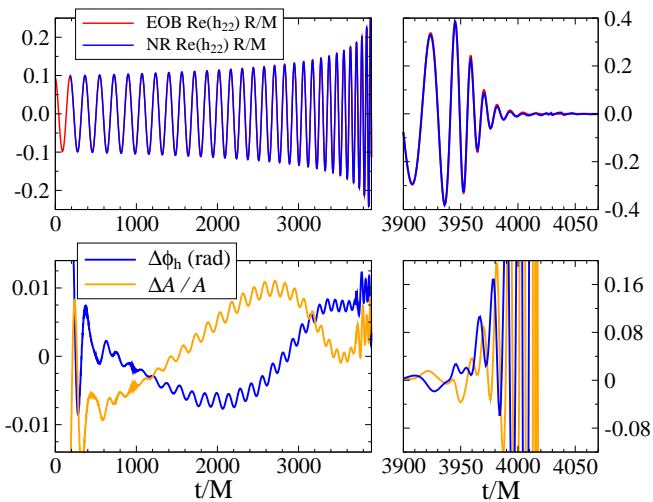}
  \caption{\label{fig:h22PhaseDiff}
Comparison of numerical waveform to EOB waveform with 
$a_5(1/4)=6.344$ and $v_{\rm pole}(1/4)=0.85$, i.e. the same model 
used in Fig.~\ref{fig:h22AmpComparison}.
The top panels show the real part of numerical and EOB
$h_{22}$, the bottom panels show amplitude and phase
  differences between them.  The left panels show
  times $t=0$ to $3900M$, and the right panels show times $t=3900$ to
  $t=4070M$ on a different vertical scale.}
\end{figure}

In Fig.~\ref{fig:h22PhaseDiff}, we show the numerical and EOB mode
$h_{22}$ aligned with the procedure of Sec.~\ref{sec:NRerrors}.  
Using the reference values $a_5(1/4)= 6.344$  and
$v_{\rm pole}(1/4)=0.85$, we find that the best phase and amplitude
agreement is obtained when the matching occurs at an interval of $\Delta t_{\rm
match}^{22}=3.0M$ ended at $t_{\rm
match}^{22}=3942.5M$, i.e., at the peak of $M\Omega$, with
$a^{h_{22}}_1(1/4)=-2.23$ and $a^{h_{22}}_2(1/4) =31.93$,
$a^{h_{22}}_3(1/4) =3.66$ and $a^{h_{22}}_4(1/4) =-10.85$.  The phase difference is strictly within
$\pm 0.02$ radians until the merger, i.e., the peak of $h_{22}$, which
happens at $t=3942.5M$ (early numerical data contaminated by junk
radiation was discarded until $t=200M$). 
The relative amplitude
difference is also within $\pm 0.02$ in this range. The phase
difference becomes $0.04$ radians at $t=3962 M$, before a rather large
error starts contaminating the numerical $h_{22}$. A more careful
tuning on the EOB-waveform adjustable parameters could further improve
the phase agreement. However, we do not think it is worthwhile to
improve the agreement at this point since we are only examining the
equal-mass case. Note that the relative amplitude difference becomes
$\sim 7\%$ at $t=3962 M$, and increases during the ringdown. 

\begin{figure}
  \includegraphics[width=\linewidth]{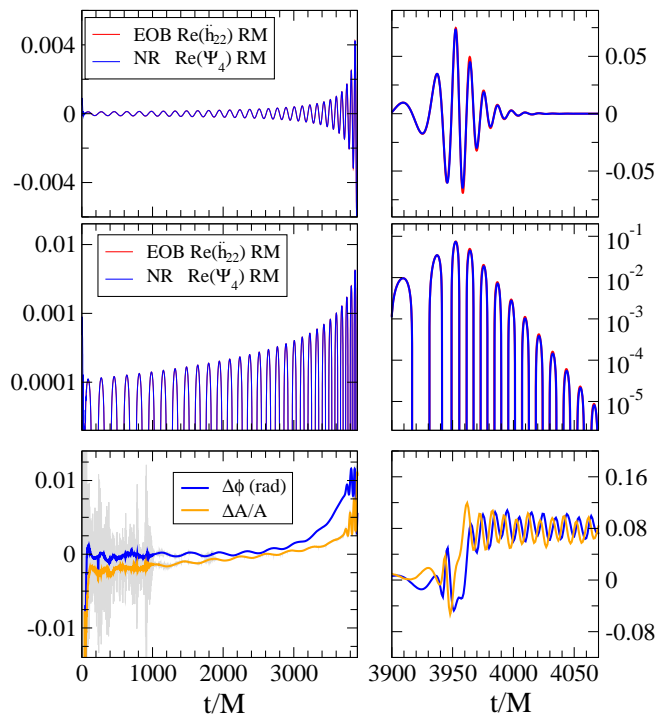}
  \caption{\label{fig:Psi422PhaseDiff} Comparison between EOB
    $\ddot{h}_{22}$ and the numerical $\Psi^{22}_4$.  The top four
    panels show the real part of the waveform, on a linear and
    logarithmic $y$-axis. The bottom two panels show the phase
    difference (in radians) and the fractional amplitude difference
    between the two waveforms.  The left panels show times $t=0$ to
    $3900M$, and the right panels show times $t=3900$ to $t=4070M$
    with different vertical scales. (The quantities in the
      lower left panel have been smoothed; the grey data in the
      background of that panel presents the raw data.)
      This figure uses the same 
EOB model as Figs.~\ref{fig:h22AmpComparison} and~\ref{fig:h22PhaseDiff}, 
namely  
$a_5(1/4)=6.344$ and $v_{\rm pole}(1/4)=0.85$.
}
\end{figure}
The numerical GW strain $h_{22}$ plotted in Fig.~\ref{fig:h22PhaseDiff} is
computed using RWZ wave extraction. 
During the ringdown, this waveform
is noisier than the extracted NP scalar
$\Psi_4^{22}$ (see the Appendix).  Therefore, in
Fig.~\ref{fig:Psi422PhaseDiff}, we compare the numerically extracted
$\Psi_4^{22}$ with the second time derivative $\ddot{h}_{22}$ of the EOB
waveform. Overall, the agreement is much better than for the comparison of
  $h_{22}$ in Fig.~\ref{fig:h22PhaseDiff}. Phase and relative
  amplitude differences are smaller than $0.002$ during most of the
  inspiral, and remain smaller than $0.01$ up to $t=3920M$. In the
  interval around merger, $t=3930M$ to $3960M$, the agreement is slightly
  worse than in Fig.~\ref{fig:h22PhaseDiff}; the disagreement in
  this region is caused by the differences between the inspiral EOB
  $\ddot{h}_{22}$ and numerical NP $\Psi_4^{22}$ frequencies, 
  as discussed at the
  beginning of Sec.~\ref{sec:EOBcalibration}.

In the ringdown region, $t>3960M$, Fig.~\ref{fig:Psi422PhaseDiff}
shows excellent agreement, and this agreement persists until late times.
In contrast to the
$h$ comparison shown in Fig.~\ref{fig:h22PhaseDiff}, 
in Fig.~\ref{fig:Psi422PhaseDiff} both phase and
amplitude differences remain {\em bounded}; during the ringdown, the
phase difference between EOB $\ddot{h}_{22}$ and $\Psi_4^{22}$ oscillates around 
$0.08$ radians, and the amplitude differs by about $8\%$. 
Apart from small oscillations likely caused by gauge effects
(see the Appendix), $\Delta\phi$ remains constant to
  an excellent degree during about 9 ringdown oscillations,
  i.e. during an accumulated phase of about 56 radians.  If the
  quasi-normal mode frequency used in the EOB ringdown waveform were
  different from the numerical ringdown frequency by as little as
  0.1\%, a linearly accumulating phase-difference of $\sim 0.056$
  radians would accumulate, which would be clearly noticeable in the
  lower right panel of Fig.~\ref{fig:Psi422PhaseDiff}.  Thus, we find
  agreement at the 0.1\% level between the numerical quasi-normal mode
  frequency and the prediction based on final mass and spin of the
  numerical simulation.

\begin{figure}
  \includegraphics[width=0.95\linewidth]{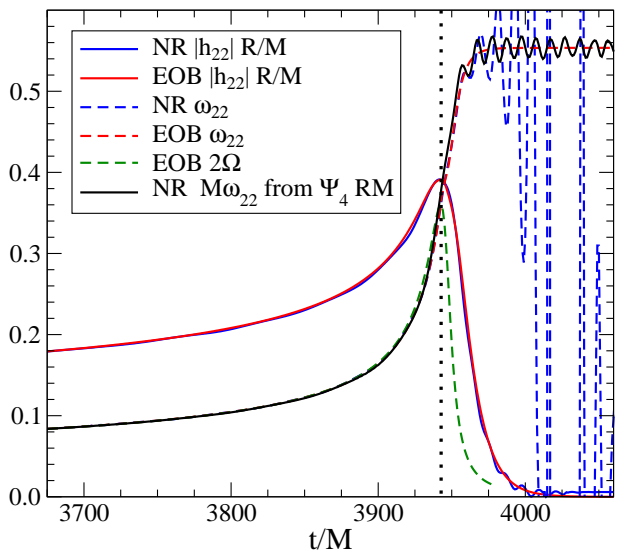}
  \caption{\label{fig:h22AmpOmega} We show the amplitude and frequency of the
numerical and EOB mode $h_{22}$, the EOB orbital frequency and 
the frequency of the numerical mode $\Psi_4^{22}$. 
The vertical line marks the peak of the EOB amplitude and 
orbital frequency.}
\end{figure}

In Fig.~\ref{fig:h22AmpOmega}, we compare the amplitude and frequency
of numerical and EOB $h_{22}$ waveforms together with the
orbital frequency of the EOB model. The peak of the latter is close to
the EOB light ring, 
and is aligned with both the EOB and numerical $h_{22}$  
amplitudes (as required by our choice of  $a^{h_{22}}_1$ and $a^{h_{22}}_2$).
During the ringdown, the frequency computed from the numerical
$h_{22}$ shows increasingly large oscillations.  We also plot the
frequency computed from the numerical $\Psi_4^{22}$ mode.
This frequency shows much smaller, and bounded, oscillations deep into
the ringdown regime.

\begin{figure}
  \includegraphics[width=0.96\linewidth]{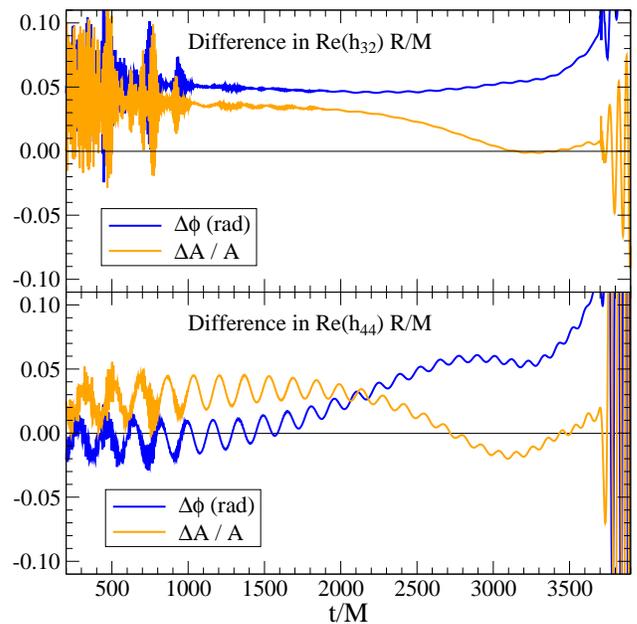}
  \caption{\label{fig:h32AmpPhaseComparison} Upper panel: 
Amplitude and phase differences of numerical and EOB mode $h_{32}$
over the inspiral range. Lower panel: Amplitude and phase
differences of numerical and EOB mode $h_{44}$ over the inspiral
range.}
\end{figure}

Having constructed our EOB waveform purely by considering the (2,2)
mode, we now discuss agreement between higher modes of the EOB
model and the numerical simulation.
Figure~\ref{fig:h32AmpPhaseComparison} shows phase and amplitude
differences for the two next largest modes, the (4,4) and the (3,2)
mode.  The EOB model is identical to the one that has been calibrated
to agree with the (2,2) mode, and the parameters $a_i^{h_{32}}$ and 
$a_i^{h_{44}}$, which appear in Eq.~(\ref{inspwavenew})
to correct the amplitude of the higher-order modes for non-quasi-circular
motion, are set to zero.
But although the EOB model has not been calibrated in any way to
match the higher-order modes, 
the agreement between numerical and EOB waveforms
shown in Figure~\ref{fig:h32AmpPhaseComparison} 
is rather good for $t\lesssim 3700M$.  In fact,
the differences between EOB and NR modes are comparable to the
estimated numerical errors in these modes (as estimated by convergence
tests between different numerical resolutions, and the comparison
between the numerical $h$ and $\Psi_4$ waveforms which are presented in
the Appendix.  Around $t\approx
3700M$, the numerical (3,2) and (4,4) modes begin to show additional
features, which we believe are unphysical, and are described in
more detail in the Appendix.  These features prevent a
meaningful comparison of the (3,2) and (4,4) modes at later times.

\begin{figure}
  \includegraphics[width=0.94\linewidth]{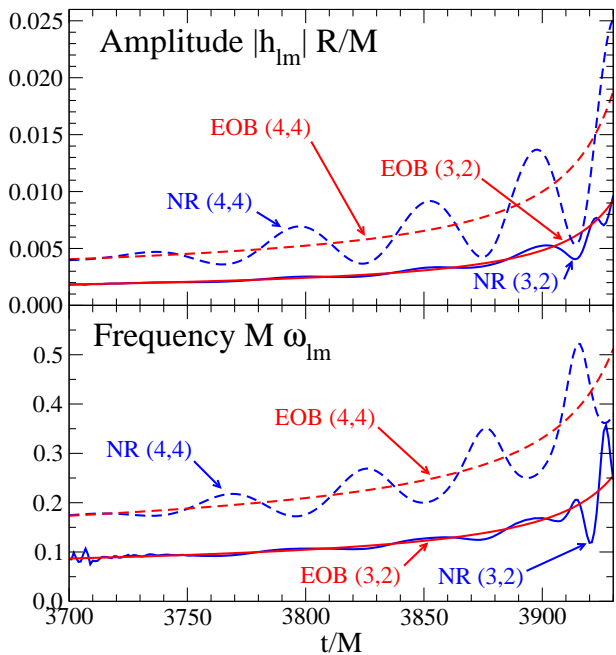}
  \caption{\label{fig:h32h44AmpOmega} Upper panel: 
Amplitude of the numerical and EOB modes $h_{32}$ and $h_{44}$.  Lower
panel: Frequency of the numerical and EOB modes $h_{32}$
and $h_{44}$. The EOB orbital frequency $2M\Omega$ ($4M\Omega$)
is indistinguishable from the frequency of the $h_{32}$ ($h_{44}$) mode
on the scale of this plot.}
\end{figure}

Figure~\ref{fig:h32h44AmpOmega} shows amplitude and frequency
  of the (4,4) and (3,2) modes for both the EOB model and the
  numerical simulation for the last few hundred M of inspiral.  This
  figure begins approximately where the NR-EOB differences in
  Fig.~\ref{fig:h32AmpPhaseComparison} exceed the vertical scale of
  that figure. The EOB amplitude and phase follow roughly the
average of numerical results, which show oscillations resulting from
numerical errors. At earlier times, the EOB and NR
  amplitudes, phase and frequencies track each other very closely, as
  can be seen from Fig.~\ref{fig:h32AmpPhaseComparison}. Please
  compare also with Fig.~\ref{fig:h22AmpOmega} which plots the
  frequencies for the (2,2) mode.

\begin{figure}
  \includegraphics[width=\linewidth]{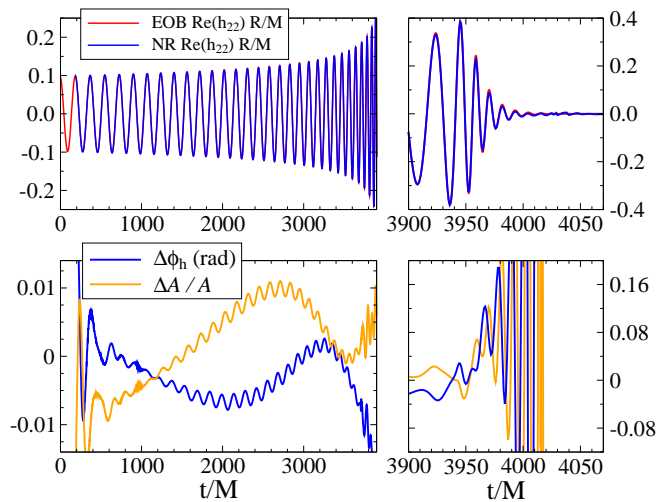}
  \caption{\label{fig:h22PhaseDiff_novp} Comparison of the numerical
    data to an EOB model with $v_{\rm pole}=\infty$.  This figure is analogous
to Fig.~\ref{fig:h22PhaseDiff}, but uses 
an EOB-model that was calibrated with the restriction $v_{\rm pole}=\infty$ (parameters are given in the main text). 
Even without $v_{\rm pole}$, the inspiral can be matched equally well as in Fig.~\ref{fig:h22PhaseDiff}; during the ringdown, the phase differences are somewhat larger, but it is possible that refined tuning will reduce them further.}
\end{figure}

Finally, in Fig.~\ref{fig:h22PhaseDiff_novp}, we show the numerical and EOB mode
$h_{22}$ using the reference values $a_5(1/4)= 4.19$ and
$v_{\rm pole} \rightarrow \infty$. In this case we find that the best phase and amplitude
agreement is obtained when the matching occurs over a range of $\Delta t_{\rm match}^{22}=2.2M$ ended at the peak of 
$M\Omega$, with $a^{h_{22}}_1(1/4)=-2.50$ and $a^{h_{22}}_2(1/4) =35.43$,
$a^{h_{22}}_3(1/4) =4.91$ and $a^{h_{22}}_4(1/4) =-32.40$.  Comparing the result with that of Fig.~\ref{fig:h22PhaseDiff}, we notice that the phase and amplitude differences are only slightly worse than
  the reference model of Fig.~\ref{fig:h22PhaseDiff}, but still within
  numerical error.

\subsection{Impact on data analysis}
\label{DA}

Using the EOB model with reference values $a_5(1/4)=6.344$ and $v_{\rm pole}(1/4)=0.85$,  
we now quantify the disagreement between numerical and EOB waveforms by calculating 
their maximized overlaps which are important for analysis  of
data~\cite{ninja} from GW detectors. Here we restrict ourselves to the dominant mode $h_{22}$.
Given two time-domain waveforms $h_1(t)$ and $h_2(t;t_0,\phi_0)$
generated with the same binary parameters, the maximized overlap, otherwise known
as a fitting factor (FF), is given explicitly by~\cite{Pan2007} 
\begin{equation}
{\rm FF}\equiv\max_{t_0,\phi_0}
\frac{\langle h_1,h_2(t_0, \phi_0)\rangle}
{\sqrt{\langle h_1,h_1\rangle\langle 
h_2(t_0, \phi_0),h_2(t_0, \phi_0)\rangle}}\,, 
\label{FF}
\end{equation}
where
\begin{equation}
\langle h_1,h_2\rangle\equiv4\,{\rm Re}
\int_0^\infty\frac{\tilde{h}_1(f)\tilde{h}^*_2(f)}{S_h(f)}\;df\,.
\label{InnerProduct}
\end{equation}
Here $\tilde{h}_i(f)$ is the Fourier transform of $h_i(t)$, and
$S_h(f)$ is the detector's power spectral density. We compute the FFs
for binary black holes with total mass $30 \mbox{--} 150 M_\odot$,
using LIGO, Enhanced LIGO and Advanced LIGO noise curves \footnote{For LIGO, we use the analytic fit to the LIGO design power spectral density given in Ref.~\cite{Damour:2000zb}; for Enhanced LIGO, we use the power spectral density given at \url{http://www.ligo.caltech.edu/~rana/NoiseData/S6/DCnoise.txt}; for Advanced LIGO, we use the broadband configuration power spectral density given at \url{http://www.ligo.caltech.edu/advLIGO/scripts/ref_des.shtml}}, and find in all cases FFs
larger than $0.999$. Note that the FFs are computed
maximizing over time of arrival and initial phase, {\it but not} over the
binary parameters. We note that ${\rm FF} \ge 0.999$ gives a mismatch 
$\epsilon \equiv {\rm 1-FF}$ between the numerical and the analytical $h_{22}$ of $\epsilon_{\rm
NR-EOB} \le 0.001$. For the noise curves of LIGO, Enhanced LIGO and Advanced LIGO,
we find that the mismatch between all extrapolated numerical waveforms $h$ 
is less than $0.0001$ for black-hole binaries with a total mass of $30 \mbox{--} 150 M_\odot$. 
If we take this mismatch as an
estimate of the difference between the numerical and the {\it exact}
physical waveforms, we have $\epsilon_{\rm e-NR} \le 0.0001$. The
mismatch between the exact and the analytical $h_{22}$ is therefore
$\epsilon_{\rm e-EOB} \le 0.0017$. This mismatch is smaller than the bound $0.005$ presented in 
Ref.~\cite{Lindblom2008}, and therefore our EOB model is sufficiently accurate
for GW detection in LIGO, Enhanced LIGO, and Advanced LIGO. 

\subsection{Unequal mass \emph{inspiraling} binary black holes}
\label{sec:comparingunequalmasses}

As a check of the robustness of our EOB model calibrated to numerical
waveforms of equal-mass black-hole binaries, we extend the model to a 
set of unequal-mass black-hole binaries by comparing numerical and EOB
$\Psi_4^{22}$ inspiraling waveforms for mass ratios 2:1 and 3:1.  These simulations were performed with 
the Caltech-Cornell SpEC code, last about eight
orbits and have phase errors similar to the equal mass simulation discussed so far.
Details of simulations will be published separately.  We explore here 
the possibility of setting $a_5(\nu) = \nu \lambda_0$ with $\lambda_0$ 
constant and let $v_{\rm pole}$ depend on the mass ratio. Indeed, in the test-particle 
limit we expect\footnote{We note that 
with our choice of the GW energy flux (factorized logarithms 
and $v_{\rm lso}=1$), 
the best fit to numerical flux has $v_{\rm pole}(0)=0.57$, 
which differs slightly} from $1/\sqrt{3}$.
$v_{\rm pole}(0)= 1/\sqrt{3} = 0.57735$, whereas in the equal-mass case we 
find $v_{\rm pole}(1/4) = 0.85$. As a preliminary study, 
we do not perform a comprehensive search over the $\lambda_0 \mbox{--}v_{\rm pole}$ parameter space
for unequal-mass binaries, as we did for equal-mass binaries in
Sec.~\ref{sec:EOBtuning}. We fix $\lambda_0$ to our reference value $25.375$
and tune $v_{\rm pole}(\nu)$ to require phase differences on the order of the 
numerical error. 

\begin{figure}
  \includegraphics[width=\linewidth]{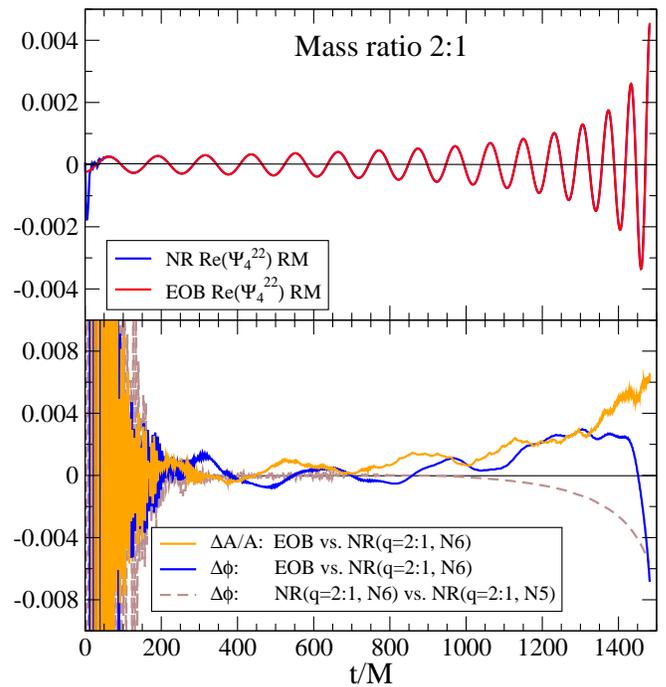}
  \caption{\label{fig:MassRatio2} {\bf EOB--NR comparison for
        a BH binary with mass ratio \boldmath $2:1$.} The upper panel
      shows the numerical and EOB mode $\Psi_4^{22}$, and the lower
      panel shows phase and amplitude differences between EOB and
      numerical run.  The dashed brown line is the estimated
      phase-error of the numerical simulation, obtained as the
      difference between simulations at high resolution 'N6' and lower
      resolution 'N5'.}
\end{figure}

\begin{figure}
  \includegraphics[width=\linewidth]{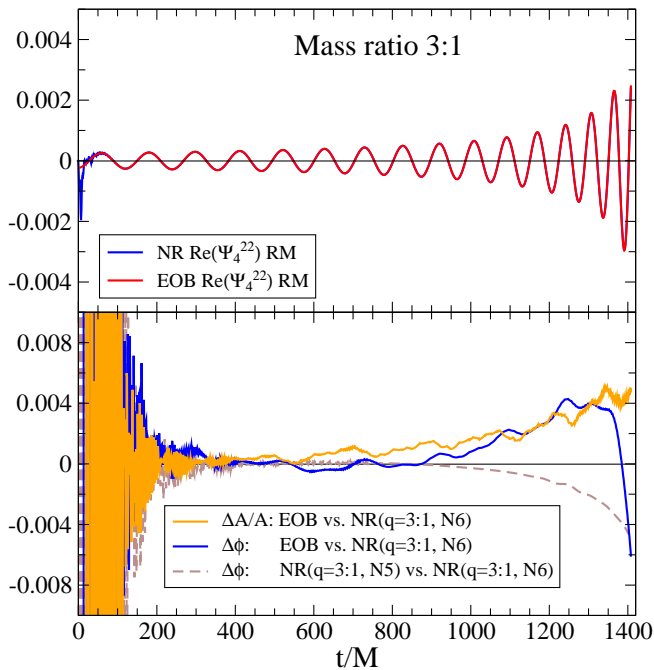}
  \caption{\label{fig:MassRatio3} {\bf EOB--NR comparison for
        a BH binary with mass ratio \boldmath $3:1$.} The upper panel
      shows the numerical and EOB mode $\Psi_4^{22}$, and the lower
      panel shows phase and amplitude differences between EOB and
      numerical run.  The dashed brown line is the estimated
      phase-error of the numerical simulation, obtained as the
      difference between simulations at high resolution 'N6' and lower
      resolution 'N5'.}
\end{figure}

In Figs.~\ref{fig:MassRatio2} and \ref{fig:MassRatio3}, we compare the
numerical and EOB $\Psi_4^{22}$ waveforms and their amplitude and
phase differences for binaries with mass ratios $q=m_1$:$m_2$ of 2:1
and 3:1.  The alignment procedure of Sec.~\ref{sec:NRerrors} was used
with $t_1=310M$ and $t_2= 930M$.  The figures also
show the numerical phase error obtained from runs with two different
numerical resolutions. The specifics of these numerical runs will be
published separately.  In the case of mass ratios $q=$ 2:1 and 3:1, we
find that by tuning $v_{\rm pole}(\nu)$, the difference between
numerical and EOB waveforms can be reduced to values smaller than the
numerical error.  The best values of $v_{\rm pole}$ we find are
$v_{\rm pole} = 0.76 \pm 0.01$ for mass ratio 2:1, and 
$v_{\rm  pole}=0.70 \pm 0.01$ for mass ratio 3:1. 
Choosing parameters outside this range results in differences between
numerical and EOB waveforms that are at least twice the numerical error.
Combining $v_{\rm pole}$ values for mass ratios 1:1, 1:2, 1:3, and the test-particle limit, we find a least-square fitting formula $v_{\rm pole}(\nu)=0.57-0.65(\pm0.35)\nu+7.0(\pm1.5)\nu^2$.

Finally, we observe that the phase and amplitude 
differences between numerical and EOB waveforms can be reduced to values 
smaller than the numerical error, if we choose the EOB reference 
model of Sec.~\ref{sec:EOBtuning} where we set $v_{\rm pole} \rightarrow \infty$ 
and let $a_5(\nu) = \nu (\lambda_0 + \lambda_1\,\nu)$. In particular, 
calibrating the mass ratio 2:1 and 3:1, we find $a_5(\nu) = \nu [-7.3 (\pm 0.1) + 
95.6 (\pm 0.3)\nu ]$.  These EOB models agree with the numerical data
as well as the EOB models shown in Figs.~\ref{fig:MassRatio2} and~\ref{fig:MassRatio3}.

\section{Conclusions}
\label{sec:conclusions}

In this paper, building upon recent, successful
results~\cite{Buonanno-Cook-Pretorius:2007,Pan2007,
Buonanno2007,Damour2007a,DN2007b,DN2008,Boyle2008a,DIN} of the EOB
formalism~\cite{Buonanno99,Buonanno00,DJS00,Damour01c,Damour03}, 
we have concentrated on the EOB model 
denoted ${}^{\rm nK}F_4^4/H_{4}$ in Ref.~\cite{Boyle2008a}, with adjustable
EOB-dynamics and EOB-waveform parameters defined in
Table~\ref{tableI}. We have calibrated this EOB model to a very accurate
numerical simulation of an equal-mass non-spinning binary black-hole 
coalescence~\cite{Scheel2008}.

When comparing EOB and numerical waveforms, or when comparing numerical
waveforms with each other, we determine the arbitrary time offset and
phase offset between the waveforms by minimizing the phase differences between
the waveforms over a time interval of $\sim 1000M$ at low frequency, where the PN-based
EOB waveforms are expected to be most accurate~\cite{Boyle2008a}.
Compared to aligning waveforms at a particular time or frequency, 
this procedure is less sensitive to numerical noise and residual
eccentricity. 

Among the EOB-dynamics adjustable parameters
$\{a_5(1/4),v_{\rm pole}(1/4), a_{\rm RR}^{{\cal F}_\Phi}(1/4),a_{\rm
RR}^{{\cal F}_r}(1/4),A_8\}$, the parameters $a_5(1/4)$ and $v_{\rm
pole}(1/4)$ have the largest effect upon the long inspiral
phase. Thus, we set $\{a_{\rm RR}^{{\cal F}_\Phi}(1/4)=0,a_{\rm
RR}^{{\cal F}_r}(1/4)=0,A_8=0\}$ in our EOB model, and we considered
the phase difference between the numerical and EOB $\Psi_4^{22}$ as a
function of the parameters $a_5(1/4)$ and $v_{\rm pole}(1/4)$.  This phase
difference increases with time, so we have sought parameters
for which this phase difference remains small for as long a time as possible.
We found regions of the $(a_5(1/4),v_{\rm pole}(1/4))$ parameter space where this
phase difference is less than $0.02$ radians either until $t=282M$ 
or until $t=42M$ (red curves) before the time when the numerical $h_{22}$ reaches
its maximum amplitude (see blue and red curves in Fig.~\ref{fig:TrefContourPlot4}), 
respectively. 

Moreover, building on Refs.~\cite{Buonanno00,BDmg9}, we have found that the EOB-dynamics adjustable
parameters entering the GW energy flux cannot modify the phase of the EOB
$\Psi_4^{22}$ during the plunge and close to merger. This
is because any modification of the GW energy flux beyond the LSO has negligible
effect on the phasing, as the evolution is driven mostly by the
conservative part of the dynamics.  We also found that $A_8$ is
strongly degenerate with $a_5(1/4)$ until almost $100M$ before merger, and that $a_{\rm
RR}^{{\cal F}_r}(1/4)$ and $a_{\rm RR}^{{\cal F}_\Phi}(1/4)$ have
an almost equivalent effect on the phasing, except for a minus sign
and a different scaling. Overall, for the equal-mass non-spinning
case, we have found that the EOB-adjustable parameters $\{a_{\rm RR}^{{\cal
F}_\Phi}(1/4),a_{\rm RR}^{{\cal F}_r}(1/4),A_8\}$ have a minor effect
in reducing the phase and amplitude differences between the EOB model
and the numerical simulation (see also 
Fig.~\ref{fig:TrefContourPlot5}). To achieve differences on the order
of the numerical error, we can restrict ourselves to the EOB parameter
space with $\{a_{\rm RR}^{{\cal F}_\Phi}(1/4)=0,a_{\rm RR}^{{\cal
F}_r}(1/4)=0,A_8=0\}$.

Furthermore, using our alignment procedure, we have found that the
peak of the numerical $h_{22}$ coincides with the peak of the EOB orbital
frequency, confirming what was pointed out in Ref.~\cite{DN2008}. 
As in Ref.~\cite{DN2008}, we require that the EOB dominant mode
$h_{22}$ peaks at the maximum of the EOB orbital frequency (i.e., the
EOB light-ring). We also require that the EOB amplitude at the peak coincides with
the numerical amplitude at the peak. In fact,
we expect that in the near future, the peak of the numerical $h_{22}$ 
will be able to be predicted by numerical relativity with high accuracy for
several mass ratios. Thus, the peak can be fit with a polynomial in $\nu$.
(Preliminary studies which use results from Ref.~\cite{Buonanno2007}
confirm this expectation.) These requirements determine the
EOB-waveform parameters $a^{h_{22}}_1(1/4)$ and $a^{h_{22}}_2(1/4)$. 
To further improved the agreement close to merger, we then tune 
$a^{h_{22}}_3(1/4)$, $a^{h_{22}}_4(1/4)$, and 
$\Delta t_{\rm match}^{22}(1/4)$, so that the phase and amplitude differences 
between the EOB and numerical $h_{22}$ are minimized. In particular, 
we found that this happens if $\Delta t_{\rm match}^{22}(1/4)$ is chosen to be around $3M$ (while $t_{\rm match}^{22}(1/4)$ is fixed at the maximum of the EOB orbital frequency $M \Omega$). For the EOB reference model with $a_5(1/4) = 6.344$ 
and $v_{\rm pole} = 0.85$, we have found that the phase and amplitude
differences between EOB and numerical $h_{22}$ waveforms
are $0.02$ radians and $2\%$, respectively, until $20 M$ 
before merger, and $0.04$ radians and
$7\%$, respectively, during merger and early ringdown, 
until the numerical
$h_{22}$ starts to be affected by numerical oscillations (see
Fig.~\ref{fig:h22PhaseDiff}).  These agreements were obtained
by comparing EOB and numerical values of $h_{22}$, the latter 
having been extracted from the RWZ scalars.  We also compared
the EOB and numerical $\Psi_4^{22}$. In this case, the agreement is
even better during the long inspiral and through the late  
ringdown, with phase
and amplitude disagreements of $0.02$ radians and $2\%$ until $20 M$ before merger, and 
$0.08$ radians and $8\%$, respectively, 
during merger and ringdown (see Fig.~\ref{fig:Psi422PhaseDiff}). 
However, around the transition 
between plunge and ringdown, the EOB $\ddot{h}_{22}$ has some oscillations because the EOB resummation provides 
us with $h_{22}$, whereas when taking time derivatives of $h_{22}$ non-resummed higher order PN 
terms are generated, spoiling in part the agreement of $\ddot{h}_{22}$.

Quite interestingly, we have found that phase and amplitude
differences between EOB and numerical waveforms can also be reduced to
numerical errors, at least during the inspiral, if we let $v_{\rm
  pole} \rightarrow \infty $ and calibrate the coefficients
$\lambda_0$ and $\lambda_1$ in $a_5(\nu) = \nu (\lambda_0 + \lambda_1
\nu)$, see Fig.~\ref{fig:h22PhaseDiff_novp}.  

For data analysis purposes, we have also computed the maximized overlaps or 
fitting factors (FFs) between the EOB reference model with $a_5(1/4) = 6.344$ 
and $v_{\rm pole} = 0.85$ and numerical $h_{22}$. We maximized only over the
initial phase and time of arrival.  We have found that for black-hole
binaries with total mass $30 \mbox{--} 150 M_\odot$, using LIGO,
Enhanced LIGO and Advanced LIGO noise curves, the FFs are larger than
0.999.  We have also computed the FFs between values of numerical $h_{22}$ 
that were computed in slightly different ways (e.g. different
numerical resolutions, different extraction procedures), and 
have estimated the mismatch between the {\it exact} and EOB $h_{22}$. We
have concluded, in the spirit of Ref.~\cite{Lindblom2008}, that our
analytical $h_{22}$ satisfies the requirements of detection with LIGO,
Enhanced LIGO and Advanced LIGO.

Finally, to test the robustness of the EOB model, we have also compared it
to a few equal-mass subdominant modes $(\ell,m)$, 
notably $(4,4)$ and $(3,2)$, and to the dominant mode $(2,2)$ of a set of unequal-mass inspiraling
binaries.  Without changing the EOB-dynamics adjustable parameters, we
have found that, in the equal-mass case, the phase and amplitude differences
of EOB and numerical $h_{44}$ and $h_{32}$ are within the numerical
errors throughout the inspiral (see
Figs.~\ref{fig:h32AmpPhaseComparison} and
\ref{fig:h32h44AmpOmega}). Furthermore, in the unequal-mass case, 
we have found that we can reduce the phase difference of the EOB
and numerical $\Psi_4^{22}$ of inspiraling binaries of mass ratios 2:1
and 3:1 on the order of the numerical error (see
Figs.~\ref{fig:MassRatio2} and \ref{fig:MassRatio3}). This can be obtained 
either (i) by setting $a_5(\nu) = \nu \lambda_0$ with $\lambda_0$ fixed by the 
equal-mass case, and calibrating $v_{\rm pole}(\nu)$, or (ii) 
by letting $v_{\rm pole} \rightarrow \infty$ and calibrating  
$a_5(\nu)$. 

In the near future, we plan to compare the non-spinning EOB model
defined in this paper to a larger set of accurate numerical
simulations of black-hole binary coalescences (for both equal and unequal-mass
binaries), and complete the tuning of all the EOB-dynamics and -waveform adjustable parameters. In
particular, we expect to improve the EOB plunge-merger-ringdown
matching either by reducing the number of EOB-waveform adjustable
parameters or by employing different matching procedures or GW energy fluxes. 

\vspace{0.1cm}
While polishing this manuscript for publication, 
an independent calibration of the EOB model which uses the equal-mass 
binary black-hole data of the Caltech-Cornell collaboration 
employed in this paper and made public on January 20, 2009 
appeared on the archives~\cite{Damour2009a}.

\begin{acknowledgments}
We thank Oliver Rinne for his work on implementing
Regge-Wheeler-Zerilli wave extraction, and Fan Zang 
for extrapolating waveforms to infinite extraction radius. We also thank 
Emanuele Berti and Evan Ochsner for useful discussions, and Emanuele Berti  
for providing us with the quasi-normal mode frequencies and decay times 
used in this paper.
  A.B. and Y.P. acknowledge support from NSF Grant No. PHY-0603762.
  L.B., L.K., H.P., and M.S. are supported in part by grants from
  the Sherman Fairchild Foundation to Caltech and Cornell, and from
  the Brinson Foundation to Caltech; by NSF Grants No. PHY-0601459,
  No. PHY-0652995, and No. DMS-0553302 at Caltech; by
  NSF Grants No. PHY-0652952, No. DMS-0553677, and No. PHY-0652929 at Cornell.
\end{acknowledgments}

\appendix*

\section{Comparing different methods of computing \boldmath$h_{\ell m}$}

The analysis in Sec.~\ref{sec:EOBcalibration} relies to some extent on the GW 
strain $h$ extracted from the numerical simulation.
Earlier papers describing generation of the numerical
data~\cite{Scheel2008,Boyle2008a,Boyle2007} focused on the behavior of
the NP scalar $\Psi_4$, and performed comparisons to PN
theory based on the numerical $\Psi_4$.  

We have two means of computing a GW strain $h$ from
the numerical simulations.  The first is a double time integration
of $\Psi_4$, exploiting the relation
\begin{equation}\label{eq:Psi4-ddoth}
\Psi_4 = \ddot{h}.
\end{equation}
[Note that throughout this Appendix, we suppress indices $\ell m$ denoting
the components of the decomposition into spin-weighted spherical
harmonics.  Thus, Eq.~(\ref{eq:Psi4-ddoth}) is meant to apply to each
complex component $(\ell,m)$].  For each time integration [and each mode
$(\ell,m)$], a complex integration constant needs to be determined.
These constants are fixed with the procedure described in Sec. II of
Ref.~\cite{Boyle2008a}, in which
a certain functional of temporal
variations of the amplitude of the integrated data is minimized.  
The minimization
is performed over 25 separate integration intervals $[t_1,t_2]$ with
$t_1/M=1000, 1100, \ldots,1400$ and $t_2/M=2600,2700, \ldots,3000$.
We then compute the time average of these 25 integrated waveforms,
and we use this time average, which
we denote as $\iint\Psi_4$, as the GW strain. Note that we
perform the above
operations on the numerical $\Psi_4$ data after it has been 
extrapolated to infinite extraction radius.  

Our second means of extracting a GW strain is using the 
RWZ equations~\cite{ReggeWheeler1957,Zerilli1970b} generalized to
arbitrary spherically symmetric coordinates, as formulated by
Sarbach~\& Tiglio~\cite{Sarbach2001}.  An advantage to the Sarbach~\&
Tiglio formalism in contrast to the more widely-used Zerilli-Moncrief
formalism (~\cite{NagarRezzolla2005} and references therein) is that
in the former case, the GW strain is obtained directly from the
gauge-invariant RWZ scalars (at leading order in the inverse radius),
without any time integration. With Oliver Rinne, we have implemented
the Sarbach~\& Tiglio formalism for a Minkowsi background in standard
coordinates in the Caltech-Cornell spectral code
\cite{Rinne2008b}. From the RWZ scalars (extracted at finite radii),
we compute the GW strain and then extrapolate to infinite extraction
radius in order to obtain the final waveform $h_{\rm RWZ}$.

\begin{figure}
\includegraphics[width=0.9\linewidth]{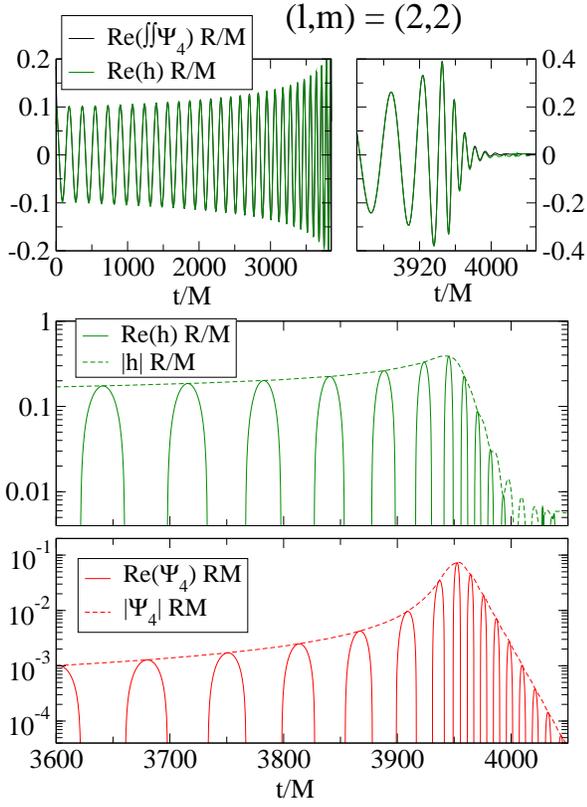}
\caption{\label{fig:WaveformRe22} The $(l,m)=(2,2)$ mode of
  the numerical waveform.}
\end{figure}
In order to gain insight into the accuracy and reliability of the
computed GW strain, we explore the differences between waveforms
extracted with either technique (see also~\cite{Pollney-Reisswig:2007} for a similar comparison).  
Figure~\ref{fig:WaveformRe22} shows
the real part of the numerical $(2,2)$ mode.  On the scale of the full
waveform, no disagreement between $h_{\rm RWZ}$ and $\iint\Psi_4$
is visible.  However, the lower two panels of
Fig.~\ref{fig:WaveformRe22} show differences between $h_{\rm RWZ}$ and
$\Psi_4$ deep in the ringdown phase: While $\Psi_4$ continues to decay
exponentially through many orders of magnitude, $h_{\rm RWZ}$ exhibits
noticeable deviation from a pure exponential decay at about a tenth of
peak amplitude.  Decay of $h_{\rm RWZ}$ stops completely at about one
per cent of peak amplitude. 

We suspect that this unexpected behavior is caused by gauge effects: 
All simulations in the numerical relativity community 
are performed using gauges in which the
coordinates dynamically respond to the changing geometry, so as to
avoid pathologies such as coordinate singularities.   Ideally, the
procedures used to extract gravitational radiation from the simulations
should be gauge invariant, so that the choice of gauge used in the simulation
is irrelevant.  In practice, however, wave extraction techniques are not
perfect.  For example, the RWZ technique is gauge invariant 
only to first order in perturbation theory about fixed background
coordinates.  Likewise, the NP technique is strictly gauge-invariant
only if applied at future null infinity, rather than at a finite distance from
the source.  
Gauge effects are expected to manifest themselves differently in
NP and RWZ wave extraction techniques, 
so by comparing the results of these
two extraction techniques, we can get a handle on the size of our uncertainties
that arise from gauge effects.

\begin{figure}
\includegraphics[width=\linewidth]{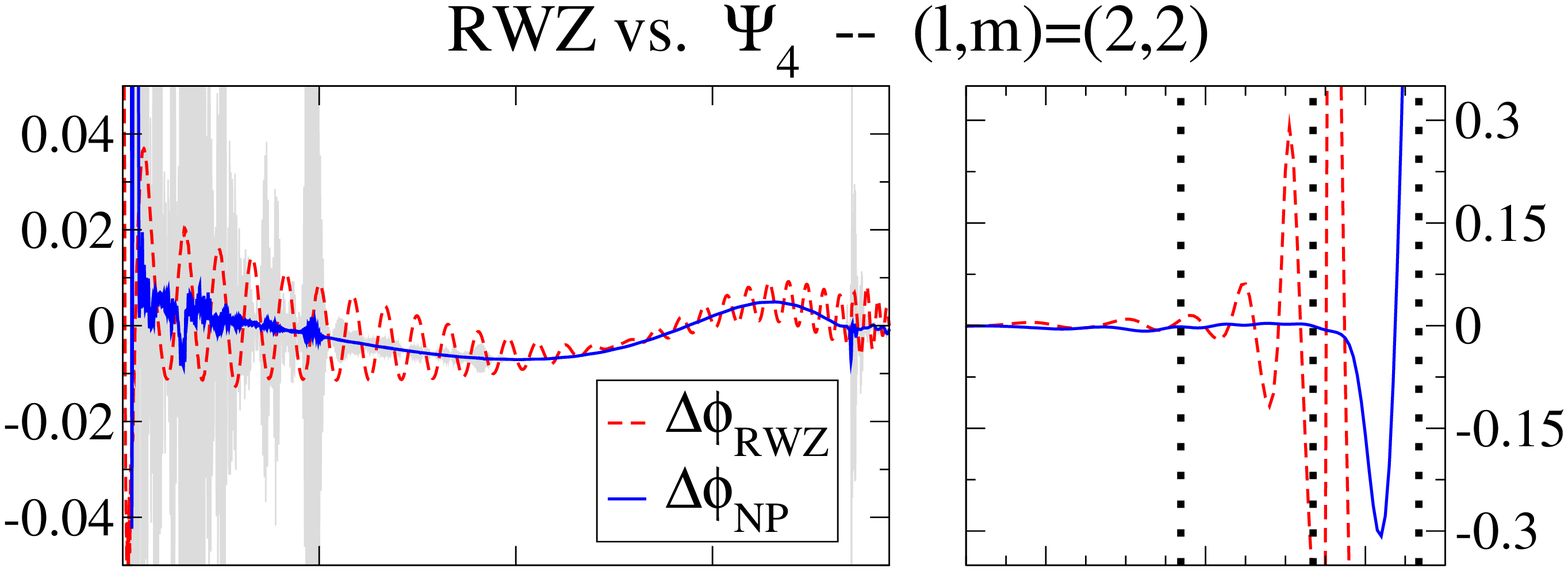}
\includegraphics[width=\linewidth]{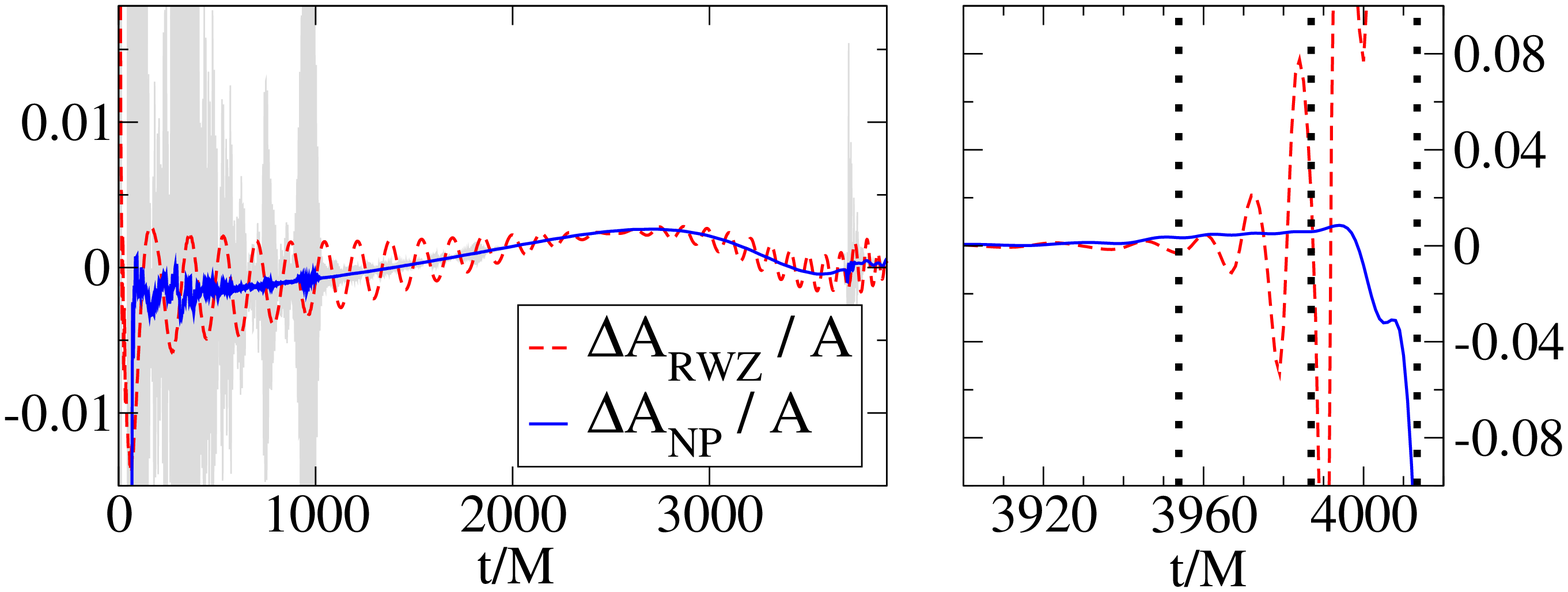}
\caption{\label{fig:RWZ-Psi4-Comparison22} 
Phase and relative amplitude
  difference between the $(l,m)\!=\!(2,2)$ modes of the RWZ waveform
  $h_{\rm RWZ}$ and NP scalar $\Psi_4$ 
[see Eqs.~(\ref{eq:DeltaPhiNP})--(\ref{eq:DeltaARWZ})]. The right
  panel shows an enlargement of merger and ringdown, with the dotted
  vertical lines indicating time of maximum of $|\Psi_4|$, and where
  $|\Psi_4|$ has decayed to 10\% and 1\% of the maximal
  value. (The blue lines are smoothed; the grey data in the background represents the unsmoothed data.)}
\end{figure}
Therefore, we will examine the differences between the numerical $h_{\rm
  RWZ}$ and $\Psi_4$.  Using (\ref{eq:Psi4-ddoth}), we can compute a
meaningful difference in two ways.  The first way is to differentiate
$h_{\rm RWZ}$ twice and compute
\begin{align}
\label{eq:DeltaPhiNP}
\Delta \phi_{\rm NP}&=\arg(\Psi_4)-\arg(\ddot h_{\rm RWZ}) \\
\frac{\Delta A_{\rm NP}}{A}&=\frac{|\Psi_4|-|\ddot h_{\rm RWZ}|}
{(|\ddot{h}_{\rm RWZ}|+|\Psi_4|)/2}.
\end{align}
The subscript 'NP' indicates that the comparison is made on the level
of the NP scalars, i.e. $\Psi_4$ appears undifferentiated
on the right-hand-sides. The second way is to time-integrate $\Psi_4$ and to calculate 
\begin{align}
\Delta \phi_{\rm RWZ}&=\arg(\mbox{$\iint\Psi_4$})-\arg( h_{\rm RWZ}) \\
\label{eq:DeltaARWZ}
\frac{\Delta A_{\rm RWZ}}{A}&=\frac{|\iint\Psi_4|-|h_{\rm RWZ}|}
{(|h_{\rm RWZ}|+|\iint\Psi_4|)/2}.
\end{align}
The results of these comparisons are presented in
Fig.~\ref{fig:RWZ-Psi4-Comparison22}.  An examination of this figure
reveals several properties of the extracted $\Psi_4$ and $h_{\rm RWZ}$
waveforms.  First, we note that during the inspiral and merger (up to
$t\lesssim 3960 M$, that is $18 M$ after the peak of $h_{\rm RWZ}$), 
the RWZ and NP waveforms agree to
better than 0.02 radians.  $\Delta\phi_{\rm NP}$ contains more noise
because noise is amplified by the double time differentiation to
compute $\ddot h$, and because $\Psi_4$ is contaminated by
junk-radiation from the initial data up to time $t\approx 1000M$.  The
blue lines in this plot have been smoothed (by convolution with a
Gaussian of width $5M$) to reduce the effect of noise due to junk
radiation.  (The grey data in the background of
Fig.~\ref{fig:RWZ-Psi4-Comparison22} shows the unsmoothed
$\Delta\Phi_{\rm NP}$).  In contrast, $\Delta\phi_{\rm RWZ}$ does not
show similar high frequency noise (the red dashed curves in
Fig.~\ref{fig:RWZ-Psi4-Comparison22} are not smoothed).  Integration
naturally smooths noise and apparently, the RWZ wave extraction is
less susceptible to the noise introduced by junk radiation.
Unfortunately, because of an imperfect choice of integration constants
for the time integration, $\iint\Psi_4$ does not precisely
oscillate around zero at all times.  This results in oscillations of
$\Delta\phi_{\rm RWZ}$ and $\Delta A_{\rm RWZ}/A$ during the inspiral;
the frequency of these oscillations coincides with the GW frequency.  
The choice of integration constants, however, is good enough to confine these
oscillations to less than about 0.02 radians in phase and 0.5 per cent
in amplitude during the inspiral.

\begin{figure}
\includegraphics[width=\linewidth]{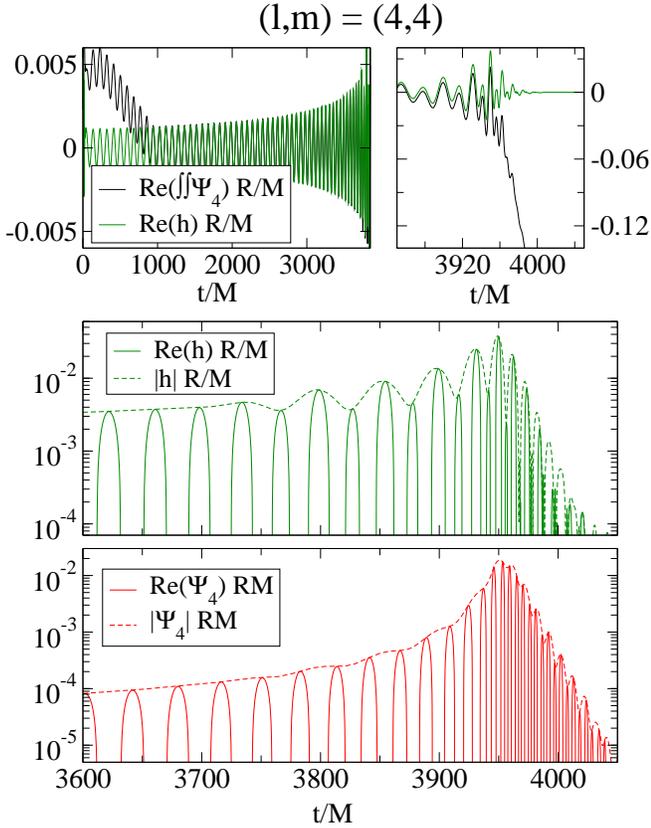}
\caption{\label{fig:WaveformRe44}The $(l,m)\!=\!(4,4)$ mode of the
  numerical waveform.}
\end{figure}

Around merger, differences of the wave strain, i.e. $\Delta\phi_{\rm
  RWZ}$ and $\Delta A_{\rm RWZ}/A$, begin to grow, and during ringdown
this growth accelerates.  This large disagreement is caused by two
effects. The first effect is the contamination of $h_{\rm RWZ}$ in the
ringdown phase, presumably by gauge effects, as shown 
in Fig.~\ref{fig:WaveformRe22}.  The
second effect is related to the time
integration used to obtain
$\iint\Psi_4$.  During the inspiral phase, with an appropriate choice of integration
constants the average value of $\iint\Psi_4$ is very nearly zero
(see top left panel of Fig.~\ref{fig:RWZ-Psi4-Comparison22}).
Thus, the inspiral phase fixes all integration constants.  When we now
extend the integration through merger and ringdown, we find that
$\iint\Psi_4$ during ringdown has a contribution that grows
linearly in time.  Because the desired oscillatory part of
$\iint\Psi_4$ decays exponentially, this linearly growing
contribution contaminates $\arg\iint\Psi_4$ to an increasing degree
as time increases.  The linearly growing contribution to
$\iint\Psi_4$ is just barely visible in the top panel of
Fig.~\ref{fig:WaveformRe22}; for the (3,2) and (4,4) modes discussed
below, it will be much more obvious. 

\begin{figure}
\includegraphics[width=\linewidth]{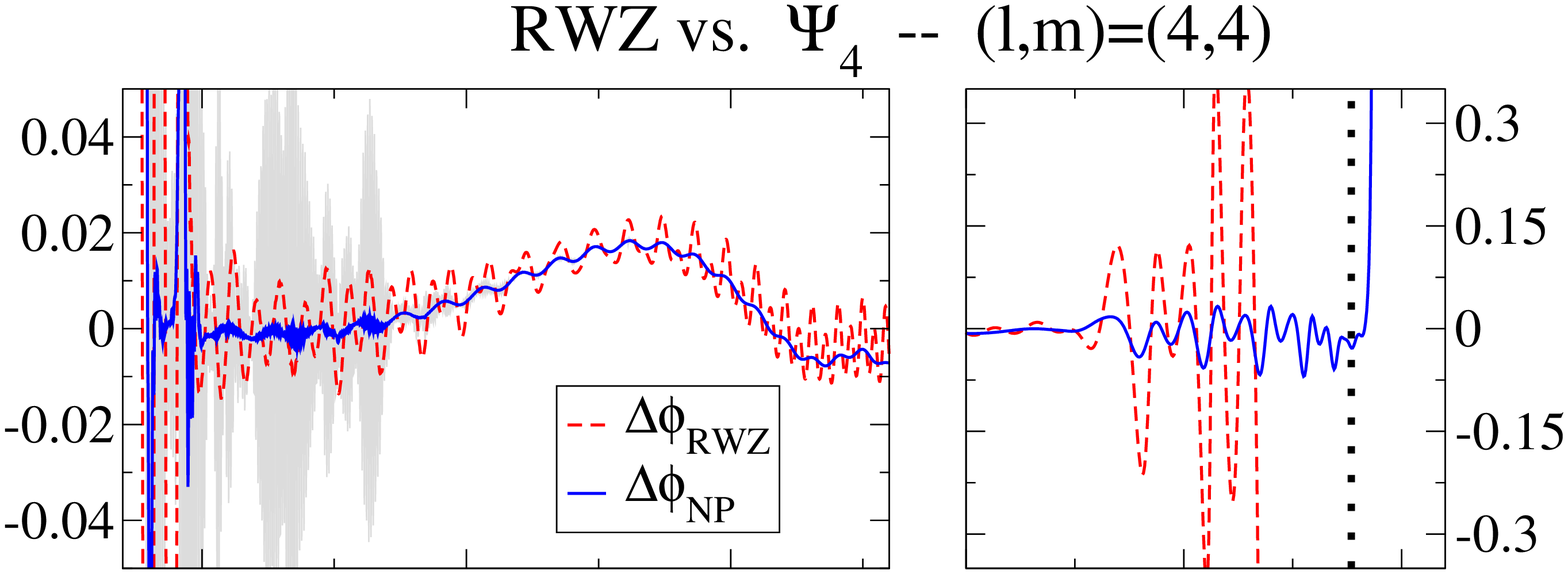}
\includegraphics[width=\linewidth]{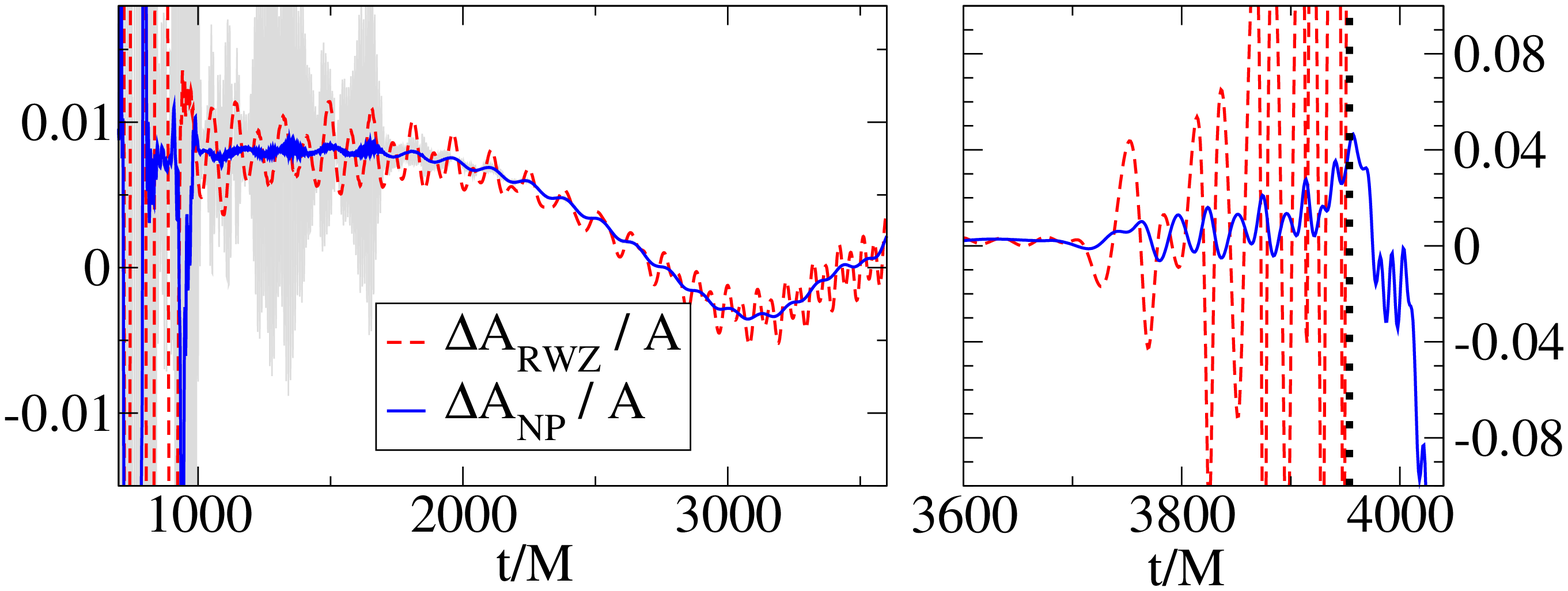}
\caption{\label{fig:RWZ-Psi4-Comparison44}Phase and relative amplitude
  difference between the $(l,m)\!=\!(4,4)$ modes of the RWZ waveform
  $h_{\rm RWZ}$ and NP scalar $\Psi_4$,
  cf. Eqs.~(\ref{eq:DeltaPhiNP})--(\ref{eq:DeltaARWZ}) The right
  panels shows an enlargement of merger and ringdown, with the dotted
  vertical line indicating the position of the maximum of
  $|\Psi_4|$. }
\end{figure}

When matching an analytical model waveform to numerical results, one
must choose whether to match to $\Psi_4$, $\iint\Psi_4$, or
$h_{\rm RWZ}$, and we have just seen that these three numerical waveforms
differ by systematic effects that arise from properties of the numerical
simulation.  Given Figs.~\ref{fig:WaveformRe22}
and~\ref{fig:RWZ-Psi4-Comparison22}, it appears that $\Psi_4$ 
is preferable over $\iint\Psi_4$ because
$\Psi_4$ lacks the low-frequency oscillations during inspiral that
are introduced in $\iint\Psi_4$ by time integration, and furthermore
$\Psi_4$ lacks the linear drift during the ringdown.  Similarly,
$\Psi_4$ has an advantage over $h_{\rm RWZ}$ because it has much cleaner
behavior during ringdown (see Fig.~\ref{fig:WaveformRe22}).

We now turn our attention to the next largest mode, $(l,m)=(4,4)$,
which is shown in
Figure~\ref{fig:WaveformRe44}. Concentrating on the top panel first, we see that 
$\iint\Psi_4$ agrees with
$h_{\rm RWZ}$ very well for a large fraction of the inspiral.
However, for $t\lesssim 1000M$ and $t\gtrsim 3900M$, $\iint\Psi_4$
contains contributions that grow linearly in time.  Note that these
contributions cannot be removed by a different choice of integration
constants, because integration constants result
in addition of a linear term
$a+bt$ {\em uniformly} at all times. Hence,
if the integration constants were changed to yield 
agreement for $t\lesssim 1000M$, the linearly growing
discrepancy would appear at $t\gtrsim 1000M$.  
The reason that the transition is around $t\sim 1000M$ may
be related to the so-called junk radiation that is present in numerical
simulations, and arises because the initial data do not
correspond precisely to a snapshot
of an evolution. A small fraction of the
outgoing junk radiation is reflected when passing through the outer
boundary.  The reflected waves pass through the computational domain
at retarded time $t\approx 1000M$.  While the reflected junk radiation
is small, apparently it is sufficient to contaminate $\iint\Psi_4$,
as seen in the top left panel of Fig.~\ref{fig:WaveformRe44}.  

\begin{figure}
\includegraphics[width=\linewidth]{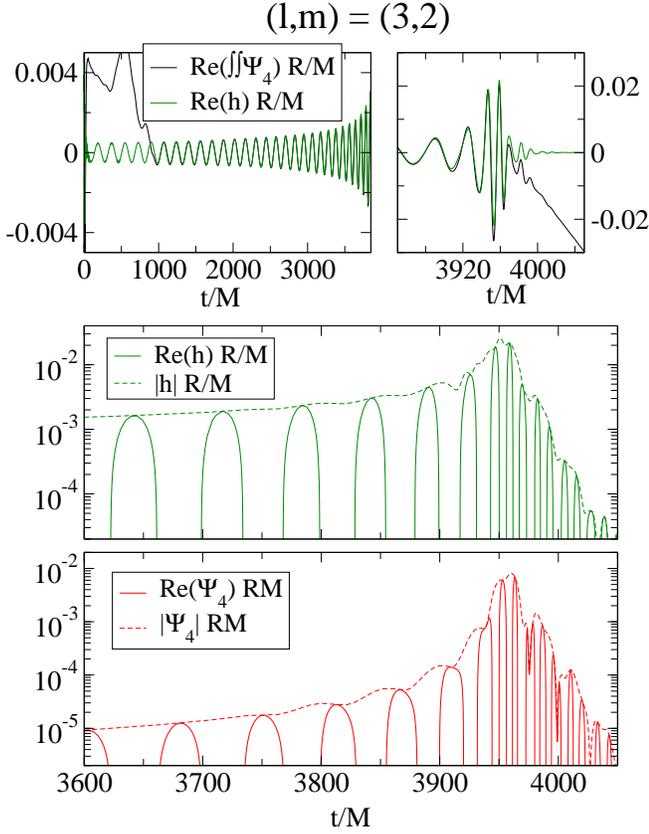}
\caption{\label{fig:WavewformRe32} The $(l,m)\!=\!(3,2)$ mode of the
  numerical waveform. }
\end{figure}

Around
merger, $t\approx 3950M$, $\iint\Psi_4$ picks up another linearly
growing contribution which renders $\iint\Psi_4$ basically useless
during merger and ringdown.  This contamination might be related to
oscillations in $\Psi_4$ and $h_{\rm RWZ}$ that become visible at
$t\gtrsim 3750M$ (see middle and lower panel of
Fig.~\ref{fig:WaveformRe44}).  It is presently unclear what causes
these effects, but we conjecture that they are related to gauge
effects that influence either our current wave-extraction procedure, or
our current wave-extrapolation procedure.  It is quite possible that a
refined understanding of gauge effects will reduce these features in
the future. 
\begin{figure}
\includegraphics[width=\linewidth]{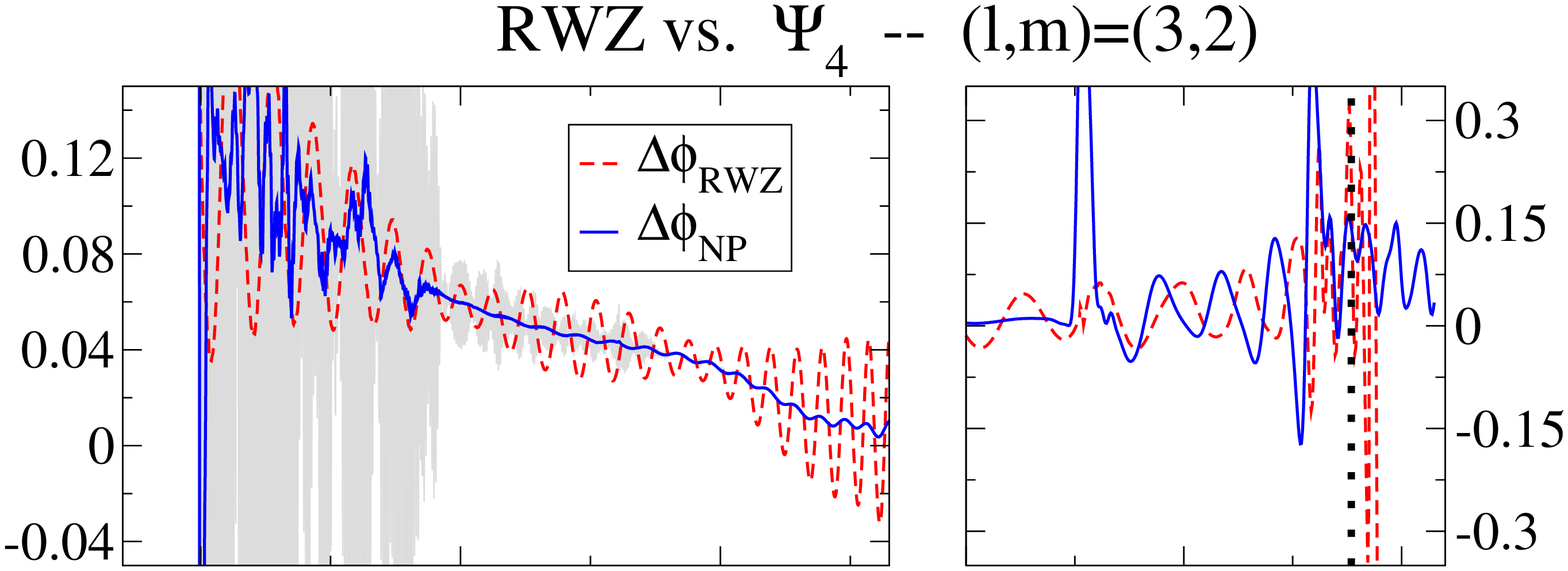}
\includegraphics[width=\linewidth]{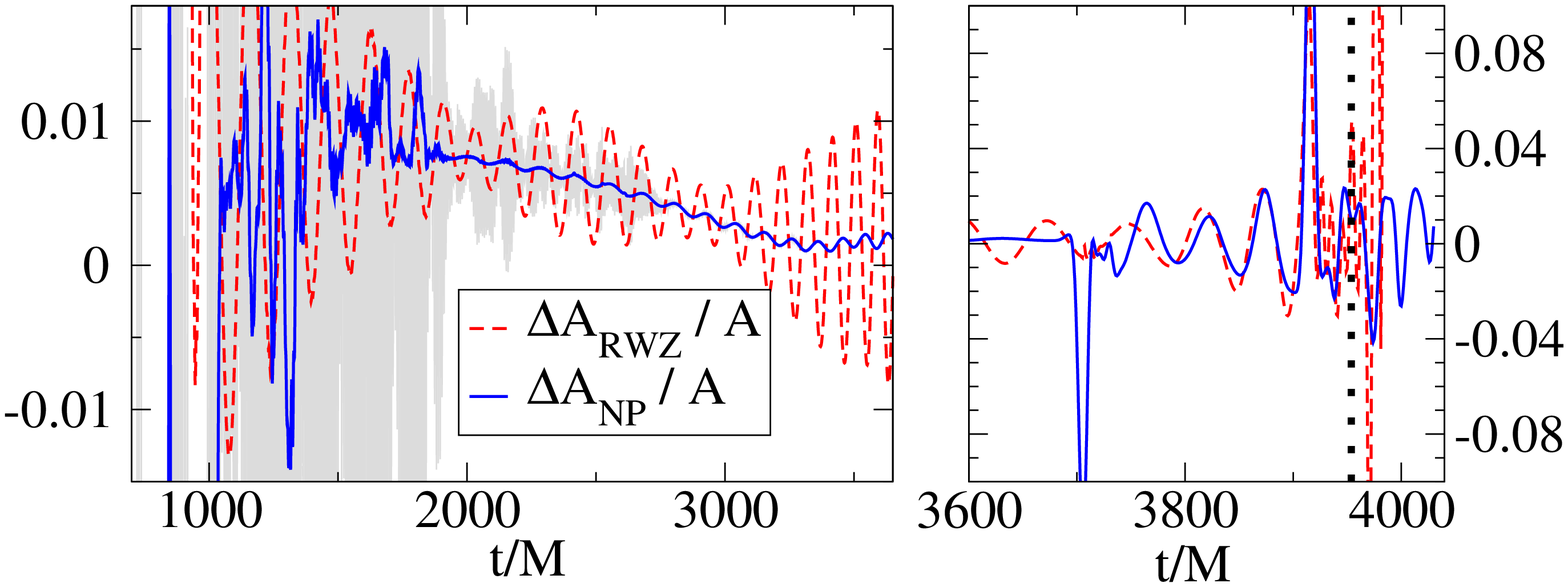}
\caption{\label{fig:RWZ-Psi4-Comparison32}Phase difference between the
  $(l,m)\!=\!(3,2)$ modes of the RWZ waveform $h$ and NP 
  scalar $\Psi_4$,
  cf. Eqs.~(\ref{eq:DeltaPhiNP})--(\ref{eq:DeltaARWZ}).  The right
  panels shows an enlargement of merger and ringdown, with the dotted
  vertical line indicating the position of the maximum of
  $|\Psi_4|$. }
\end{figure}

Because of the apparent contamination of the waveforms for early and
late times, we will restrict the EOB-NR comparison of the higher order
modes to the time interval $1000\lesssim t/M\lesssim 3600$.
Figure~\ref{fig:RWZ-Psi4-Comparison44} shows that within this
interval,$\Psi_4$ and $h_{\rm RWZ}$ agree to better than $0.02$ radians
in phase and $1\%$ in amplitude.  

Finally, Figures~\ref{fig:WavewformRe32} and
\ref{fig:RWZ-Psi4-Comparison32} present an analogous comparison for
the $(l,m)=(3,2)$ mode.  Qualitatively, these figures are similar to
Figs.~\ref{fig:WaveformRe44}
and~\ref{fig:RWZ-Psi4-Comparison44}. Agreement between $\Psi_4$ and
$h_{\rm RWZ}$ is very good for the time interval $1000\lesssim
t/M\lesssim 3700M$, with the phases differing by less than $0.1$ radians
and the amplitudes by less than about $1\%$.  The larger
disagreement might be due to the smaller amplitude of the (3,2) mode
of $\Psi_4$ during the inspiral phase relative to the (4,4) mode.  One
potentially interesting difference between the (4,4) and (3,2) modes
lies in the relative size of the variations in $|h_{\rm RWZ}|$ and
$|\Psi_4|$ in the time range $3700\lesssim t/M\lesssim 3900M$: For
the (4,4) mode, variations in $|\Psi_4|$ are clearly smaller than
variations in $|h_{\rm RWZ}|$ (see Fig.~\ref{fig:WaveformRe44}.  For
the (3,2) mode this is reversed, with $|h_{\rm RWZ}|$ showing 
somewhat smaller variations than $|\Psi_4|$.

\bibliography{References}
\end{document}